\documentclass[pra,aps,reprint,a4paper,superscriptaddress,longbibliography,floatfix]{revtex4-1}
\usepackage{times}
\usepackage{epsfig}
\usepackage{amsfonts}
\usepackage{amsmath}
\usepackage{amssymb}
\usepackage{color}
\usepackage{multirow}

\usepackage[normalem]{ulem}
\usepackage{latexsym}
\usepackage{amsfonts}
\usepackage{mathrsfs}
\usepackage{natbib}
\usepackage{verbatim}
\usepackage{amsmath}
\usepackage{amsfonts}
\usepackage{amssymb}
\usepackage{graphicx}
\usepackage[colorlinks=true,linkcolor=blue,citecolor=magenta,urlcolor=blue]{hyperref}
\newcommand{\la}{\langle}
\newcommand{\ra}{\rangle}
\usepackage{natbib}
\usepackage{color}

\usepackage[colorlinks=true,linkcolor=blue,citecolor=magenta,urlcolor=blue]{hyperref}

\DeclareMathOperator{\Tr}{tr}

\newcommand{\ket}[1]{|#1\rangle}

\begin{document}

%%%%%%%%%%%%%%%%%%%%%%%%%%%%%%%%%%%%%%%%%%%%%%%%%%%%%%%%%%%%%%%%%%%
\title{High-dimensional quantum communication complexity beyond strategies based on Bell's theorem}
%\title{Experimental observation of dimensional discontinuity in quantum communication complexity}
%\title{Experimental observation of dimensional threshold for better-than-entanglement-based quantum communication complexity}

%%%%%%%%%%%%%%%%%%%%%%%%%%%%%%%%%%%%%%%%%%%%%%%%%%%%%%%%%%%%%%%%%%%

\author{Daniel Mart\'inez}
\affiliation{Departamento de F\'{\i}sica, Universidad de Concepci\'on, 160-C Concepci\'on, Chile}
\affiliation{Millennium Institute for Research in Optics, Universidad de Concepci\'on, 160-C Concepci\'on, Chile}

\author{Armin Tavakoli}
\affiliation{Groupe de Physique Appliqu\'ee, Universit\'e de Gen\`eve, CH-1211 Gen\`eve, Switzerland}

\author{Mauricio Casanova}
\affiliation{Departamento de F\'{\i}sica, Universidad de Concepci\'on, 160-C Concepci\'on, Chile}
\affiliation{Millennium Institute for Research in Optics, Universidad de Concepci\'on, 160-C Concepci\'on, Chile}

\author{Gustavo Ca{\~n}as}
\affiliation{Departamento de F\'{\i}sica, Universidad del Bio-Bio, Avenida Collao 1202, Concepci\'on, Chile}

\author{Breno Marques}
\affiliation{Instituto de F\'isica, Universidade de S\~ao Paulo, 05315-970 S\~ao Paulo, Brazil}

\author{Gustavo Lima}
\affiliation{Departamento de F\'{\i}sica, Universidad de Concepci\'on, 160-C Concepci\'on, Chile}
\affiliation{Millennium Institute for Research in Optics, Universidad de Concepci\'on, 160-C Concepci\'on, Chile}

%%%%%%%%%%%%%%%%%%%%%%%%%%%%%%%%%%%%%%%%%%%%%%%%%%%%%%%%%%%%%%%%%%%
\begin{abstract}
Quantum resources can improve communication complexity problems (CCPs) beyond their classical constraints. One quantum approach is to share entanglement and create correlations violating a Bell inequality, which can then assist classical communication. A second approach is to resort solely to the preparation, transmission and measurement of a single quantum system; in other words quantum communication. Here, we show the advantages of the latter over the former in high-dimensional Hilbert space. We focus on a family of CCPs, based on facet Bell inequalities, study the advantage of high-dimensional quantum communication, and realise such quantum communication strategies using up to ten-dimensional systems. The experiment demonstrates, for growing dimension, an increasing advantage over quantum strategies based on Bell inequality violation. For sufficiently high dimensions, quantum communication also surpasses the limitations of the post-quantum Bell correlations obeying only locality in the macroscopic limit. Surprisingly, we find that the advantages are tied to the use of measurements that are not rank-one projective. We provide an experimental semi-device-independent falsification of such measurements in Hilbert space dimension six.
\end{abstract}
%%%%%%%%%%%%%%%%%%%%%%%%%%%%%%%%%%%%%%%%%%%%%%%%%%%%%%%%%%%%%%%%%%%

\maketitle

%%%%%%%%%%%%%%%%%%%%%%%%%%%%%%%%%%%%%%%%%%%%%%%%%%%%%%%%%%%%%%%%%%%

\textit{Introduction.---} Communication complexity problems (CCPs) are tasks in which distant parties hold local data, the collection of which is needed for a computation of their interest.~To make the computation possible, the parties communicate with each other.~However, the amount of communication is limited and therefore not all data can be sent. The CCP consist in parties adopting an efficient communication strategy which allows them to perform the desired computation with a probability as high as possible. Efficient use of quantitatively limited communication is a broadly relevant matter \cite{KSbook}, which provides fundamental insights on physical limitations \cite{BB06, IC09}.

The ability to process information depends on the choice of the physical system into which the information is encoded \cite{Landauer}. Consequently, quantum entities without a classical counterpart can be regarded as tools for quantum information processing. The most famous example is entanglement. In a quantum CCP, parties may share an entangled state on which they perform local measurements, generating strongly correlated data which violates a Bell inequality. That data can then be used to assist a classical communication strategy \cite{ComplexityReview}. In fact, Bell inequalities have been systematically linked to CCPs \cite{BZ04, BC16, TZ17}, and their violation enables better-than-classical communication efficiencies  \cite{CB97, Buhrman2,BZZ02, BP03, BZ04, EB13, HD199, Bridge, TZ17}.

Nevertheless, quantum theory presents also a second approach to CCPs:~substituting classical communication with quantum communication. The justification for such a substitution relies on the Holevo theorem \cite{Holevo} which implies that no more information can be extracted from quantum $d$-level system than from a classical $d$-level system. Hence, in a quantum communication strategy, information is encoded in a quantum state of a specified Hilbert space dimension, and subsequently extracted by a measurement. The ability of quantum communication to outperform classical constraints in CCPs is well-established \cite{RAC1, RACthesis, WignerFunction, G02, Nayak, TS05, GF16, TH15, SE16}.

Many quantum communication tasks can be successfully completed both by means of local measurements on an entangled state followed by classical communication, or by the communication of a single quantum system \cite{Crypto, Byzantine, Secretsharing}. For two-party CCPs with binary communication followed by binary-outcome measurements, classical communication assisted by correlations violating a Bell inequality is always at least as good as an implementation based on quantum communication \cite{PW}. Explicit examples in which the advantage is strict are known \cite{PZ10, HS17}. However, there also exists examples of particular scenarios of two-party CCPs with more than two outcomes in which quantum communication holds an advantage over the Bell inequality based approach \cite{TM16, magic7}.

In this work, we theoretically explore and experimentally demonstrate advantages of performing CCPs with quantum communication in high-dimensional Hilbert space, as compared to exploiting the violation of a Bell inequality. To this end, we focus on a family of CCPs \cite{BZZ02, BP03} based on the (to the best of our knowledge) only known family of bipartite facet Bell inequalities, namely the Collins-Gisin-Linden-Massar-Popescu (CGLMP) inequalities, involving any $d$ number of outcomes \cite{cglmp02, M02}. We demonstrate the advantage of quantum communication over strategies based on violations of the CGLMP inequalities, which we show to be even larger than previously thought \cite{magic7}. In particular, whilst resolving two conjectures of \cite{magic7}, we show that below dimension six, both quantum CCP-implementations are equally efficient, whereas above (and including) dimension six they are not. In this sense, dimension six acts as a threshold for revealing the advantages of quantum communication. To shine light on the suddenly emerging discrepancy between the two quantum CCP-implementations, we evidence that optimal quantum communication strategies in high-dimensional Hilbert space require projective measurements that are not rank-one. Subsequently, we present an experimental realisation. Using high-dimensional photonic systems, specifically up to dimension ten, we outperform strategies based on violating the CGLMP inequalities, emerging from dimension six, by means of quantum nonlocal correlations. Furthermore, we also outperform strategies based on super-quantum violations of said inequalities respecting only no-signaling and macroscopic locality \cite{ML}.  Finally, we prove that the experimental data cannot be simulated with any rank-one projective measurement without additional post-processing of the data. Since only a dimensional bound on the relevant Hilbert space is assumed, this constitutes a semi-device-independent \cite{PB11} falsification of said property.

\textit{The communication complexity problems.---} Bell inequalities can be systematically mapped to CCPs. In a Bell experiment, any choice of shared state and local measurements, which then generates a probability distribution, can also be used in a strategy for a CCP leading to an efficiency analogous to that observed in the Bell experiment \cite{BZ04, TZ17}. A quantum advantage (in such strategies) over classical methods  relies on generating correlations that violate the relevant Bell inequality.  A natural candidate for such constructions are facet Bell inequalities, since these optimally bound correlations obeying local realism. The CGLMP inequalities \cite{cglmp02} constitute a family of facet Bell inequalities for two parties,  each with two choices of measurements and with $d\geq 2$ possible outcomes.

The construction of CCPs based on the CGLMP inequalities has been developed in \cite{BZZ02, BP03}. In this family of CCPs (parameterised by $d$), a party Alice is given random inputs $x_0\in\{0,\ldots,d-1\}$ and $x\in\{0,1\}$, and another party, Bob, is given a random input $y\in\{0,1\}$. Alice may communicate no more than $\log d$ bits to Bob, after which he outputs a guess $g\in\{0,\ldots,d-1\}$. If $g$ coincides with the value of a function $f_k(x_0,x,y)=x_0-xy-(-1)^{x+y}k\mod{d}$, for some $k=0,\ldots,\lfloor d/2\rfloor -1$, the partnership is awarded $c_k=1-2k/(d-1)$ points. However, if $g$ coincides with $h_k=x_0-xy+(-1)^{x+y}(k+1)\mod{d}$, the partnership loses $c_k$ points. The task is to efficiently communicate such that the average number of points earned is large. The payoff function is given by
\begin{eqnarray}
\nonumber \Delta_d^{\text{Bell}}&=&\frac{1}{4d}\sum_{\substack{x_0,x\\y,k}}c_k\left[P(g=f_k|x_0,x,y)-P(g=h_k|x_0,x,y)\right]. \nonumber
\end{eqnarray}

\begin{figure}
	\centering
	\includegraphics[width=0.85\columnwidth]{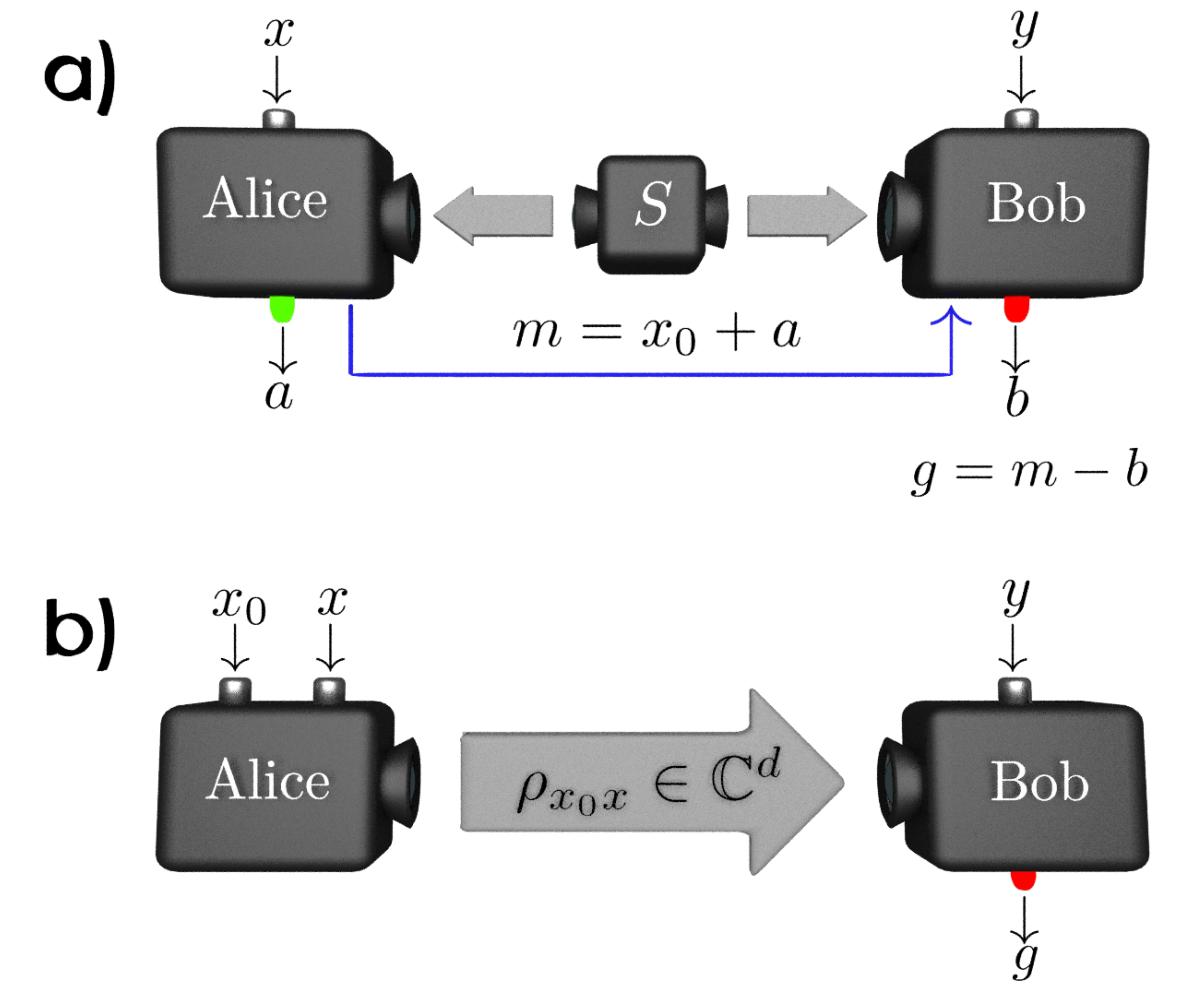}
	\caption{(a) Quantum CCP implementation based on the violation of the CGLMP inequalities. (b) Quantum CCP implementation based on communicating a single $d$-level quantum system.}\label{figscen}
\end{figure}

On the one hand, in an approach based on Bell inequalities, Alice and Bob share an entangled state and perform local measurements $x$ and $y$ with $d$-valued outcomes $a$ and $b$ respectively. In order to exploit the fact that the CCP is tailored to the CGLMP inequalities, Alice encodes the classical communication $m(a,x_0,x)\in\{0,\ldots,d-1\}$ using $m=x_0+a\mod{d}$ and Bob subsequently decodes it using $g=m-b\mod{d}$ (see Fig.~\ref{figscen}).
 It was shown \cite{BZZ02, BP03, magic7} that the resulting value of $\Delta_d^{\text{Bell}}$ is in one-to-one correspondence with the quantity evaluated from the statistics  $p(a,b|x,y)$ in a test of the CGLMP inequalities. In this sense, the efficiency in the CCP is determined by the amount of nonlocality  present in the distribution $p(a,b|x,y)$. In particular, if $p(a,b|x,y)$ generates a maximal violation of the (suitably normalised) CGLMP inequalities, then by the outlined communication strategy it can be used to achieve an equally large value of $\Delta_d^{\text{Bell}}$. The maximal quantum value achievable in a test of the CGLMP inequalities lacks a simple analytical form but is known up to large $d$ and achieved with non-maximally entanged states \cite{ZG08}.

On the other hand, these CCPs can also be implemented without exploiting entanglement and Bell inequality violations \cite{magic7}. Instead, Alice and Bob can use single quantum systems for direct quantum communication. In such an implementation, Alice associates her random inputs $(x,x_0)$ to a $d$-dimensional quantum state, $\rho_{x_0x}\in \mathbb{C}^d$, which is sent to Bob who performs a measurement $\{M_y^{g}\}_{g=0}^{d-1}$, the outcome $g$ of which determines his output guess (see Fig.~\ref{figscen}). In a quantum model, the performance of the CCP reads
\begin{equation}
\Delta_d^{\text{QS}}=\frac{1}{4d}\sum_{x_0,x,y,k}c_k\Tr\left(\rho_{x_0x}\left(M_y^{f_k}-M_y^{h_k}\right)\right).
\end{equation} An efficient quantum communication strategy, i.e., a suitable choice of state preparations and measurements, aims to find the largest value of $\Delta_d^{\text{QS}}$. In  Ref.~\cite{magic7}, it was shown that the optimal performance of the two different quantum approaches is equal, i.e., $\Delta_d^{\text{QS}}=\Delta_d^{\text{Bell}}$, when $d=2,3,4$. Numerical results suggested the same relation also for $d=5,6$. However, for $d\geq 7$, lower bounds on $\Delta_d^{\text{QS}}$ were shown to outperform the maximal value of $\Delta_d^{\text{Bell}}$. Next, we revisit this analysis, show improved quantitative results, establish the precise dimension revealing the inequivalence, and provide insight to the qualitative differences between the two quantum implementations of the CCPs.

\textit{The efficiency of quantum communication.---} We begin by quantifying the advantage of quantum communication  over strategies based on the violation of the CGLMP inequalities. Specifically, we numerically infer lower bounds on the maximal value of $\Delta_d^{\text{QS}}$ for $d\leq 10$. This has been done by running two optimisations in see-saw \cite{Seesaw1, Seesaw2}; first optimising over the states of Alice for fix measurements of Bob, and then over the measurements of Bob for fix states of Alice, repeatedly. Each such optimisation constitutes a  semidefinite program \cite{sdp}. The best found states and measurements are listed in Appendix.~\ref{App}. The results are presented in Table~\ref{Tab1} together with the known \cite{ZG08, magic7} optimal CGLMP-based values of $\Delta_d^{\text{Bell}}$ as obtained both in quantum theory, and by the super-quantum principle of macroscopic locality \cite{ML}. The latter correlations are only constrained by the inability of violating a Bell inequality when the measurements are macroscopic, i.e., that a large number of particles are collectively measured instead of microscopic measurements on single particles. The results substantially improve on the lower bounds for $\Delta_d^{\text{QS}}$ obtained in \cite{magic7}, and thus establish an increased quantitative advantage of high-dimensional quantum communication over strategies based on Bell inequality violation. In particular, note that for $d=8,9,10$, quantum communication can even outperform the Bell inequality based approach when the correlations established are only required to be macroscopically local, i.e., the violation of the CGLMP inequalities is larger-than-quantum.

\begin{table}[t]
	\centering
	\begin{tabular}{|c|c|c|c|c|c|c|}
		\hline
		$d$ & $ \substack{\text{Lower bound}\\ \Delta^{\text{QS}}_d} $ & $ \substack{\text{Lower bound }\\  \Delta^{\text{QS}}_d \text{ from \cite{magic7}}}  $ & $\Delta^{\text{Bell}}_d$ &  $\Delta_d^{\text{ML}}$  &  $\substack{\text{Lower bound} \Delta_d^{\text{QS}}\\ \text{rank-one projective}}$\\ [0.5ex]
		\hline
		2 & -  & 0.7071 & 0.7071 & 0.7071 & 0.7071 \\
		3 & - & 0.7287 & 0.7287 & 0.7887 & 0.7287\\
		4 & -  & 0.7432 & 0.7432& 0.8032 & 0.7432	\\
		5 & - &  0.7539 & 0.7539 & 0.8249 & 0.7539\\
		6 & 0.8000 & 0.7624 & 0.7624 & 0.8345 & 0.7624  \\
		7 & 0.8175  & 0.7815 & 0.7694 & 0.8461 & 0.7814 \\
		8 & 0.8571  & 0.8006 & 0.7753 & 0.8529 & 0.8006\\
		9 & 0.8622  & 0.8622 & 0.7804 & 0.8605 & 0.8188\\
		10 & 0.8889 & 0.8778 & 0.7849 & 0.8657 & 0.8396\\
		\hline
	\end{tabular}
	\caption{Lower bounds for the maximal value of $\Delta_d^{\text{QS}}$ as compared to the maximal value of $\Delta_d^{\text{Bell}}$ obtained via the maximal quantum (and macroscopically local i.e., $\Delta_d^{\text{ML}}$) violation of the CGLMP inequalities. The final column was obtained through optimisation over unit-trace measurement operators and optimal measurements were always found to be rank-one projective.}
	\label{Tab1}
\end{table}

Furthermore, we rectify the main result of \cite{magic7} by resolving two of its conjectures:  that the optimal quantum communication strategy performs equally well as that based on the quantum violation of the CGLMP inequalities when $d=5$ and when $d=6$. For $d=5$, we have used the second hierarchy level of dimensionally bounded quantum correlations \cite{NV}. In order to reduce the computational requirements of this evaluation,  we have employed the symmetrisation techniques and toolbox of \cite{symmetrisation}. We obtain a tight bound on the efficiency of quantum communication matching that obtained through a maximal violation of the CGLMP inequalities. This proves the conjecture. For $d=6$, the presented lower bound on $\Delta_d^{\text{QS}}$ shows that quantum communication outperforms the analogous Bell inequality based result. Thus, the improved lower bound falsifies the conjecture. This establishes dimension six as the dimension revealing the quantitative inequivalence between the two quantum implementations of the CCPs.

A relevant question is whether the breaking of the equivalence of the two quantum implementations,  emerging when the dimension is increased above five, is linked to qualitatively different properties in the optimal use of the respective quantum systems.   Below the critical dimension six, the optimal value of $\Delta_d^{\text{QS}}$ is obtained by Alice's average ensemble hiding the value of $x$ (in the spirit of no-signaling) and is equivalent to the post-measurement states of Bob in an optimal strategy based on the CGLMP inequalities. Via known analogies between the quantum strategies for the two cases  \cite{TZ17}, it is optimal to perform the same pair of rank-one projective measurements on Bob's side. However, when $d\geq 6$ our numerical calculations for $d=6,\ldots, 10$ suggests that: (I) the states $\{\rho_{x_0x}\}$ sent by Alice do not emulate the no-signaling condition (i.e., do not hide the value of $x$) and some inputs may be associated to the same state, and (II) the two measurements of Bob are such that one is rank-one projective, whereas the other is higher-rank projective, i.e., some measurement operators are zero-operators, meaning that the associated outcomes never can occur regardless of the state being measured. These latter measurements can be viewed as rank-one projective measurements with additional post-processing by which some outcomes remain untouched and other outcomes are given new labels. To further evidence the sub-optimality of rank-one projective measurements (without post-processing), we have  numerically optimised $\Delta_d^{\text{QS}}$ over measurements in which all measurement operators are of trace one. Since all rank-one projectors are of trace one, and we always find the optimal measurement to be rank-one projective, the results constitutes a lower bound valid for such measurements. The results (see Table~\ref{Tab1}) show that although rank-one projective measurements are sufficient to outperform strategies based violating the CGLMP inequalities, they are not optimal.

\textit{Experimental demonstration of high-dimensional quantum communication advantage.---}
\begin{figure}
	\centering
	\includegraphics[width=\columnwidth]{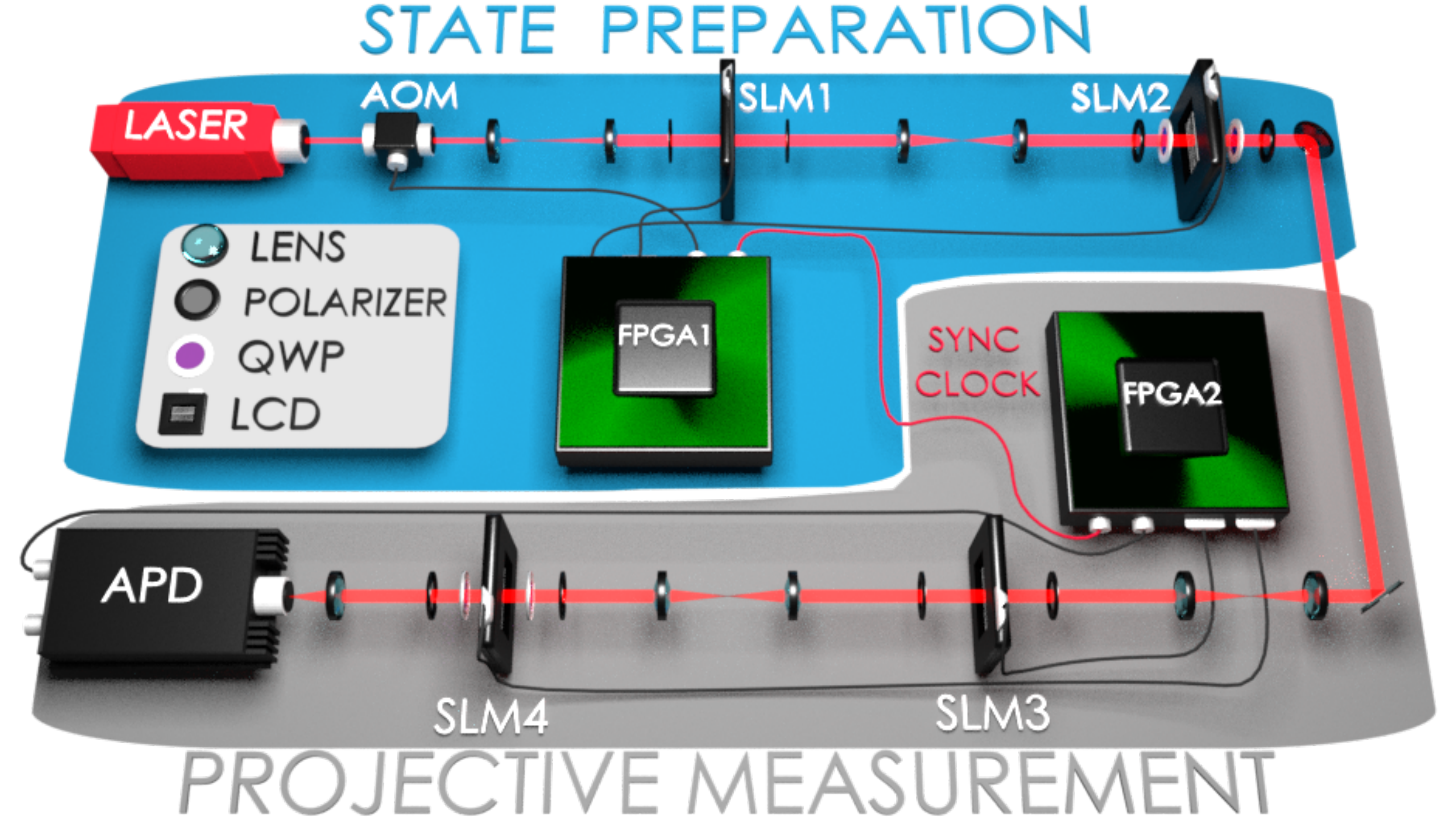}
	\caption{Experimental setup for implementing the CCPs with quantum communication. $d$-dimensional quantum systems are encoded into the linear transverse momentum of single photons. The experiment is composed of two main parts: one for the state preparation and another for performing measurements on the prepared system. Both parts rely on the programmability of spatial light modulators for preparing the required states and measurements.}
	\label{fig:setup}
\end{figure}
We present an experimental demonstration of the advantages of single-system quantum communication in the considered CCPs for $d = 6,...,10$. In our experiment, $d$-dimensional quantum systems are encoded into the linear transverse momentum of single photons transmitted by programmable diffractive apertures, which nowadays is a standard technique used for high-dimensional quantum information processing \cite{GLima03,TomoMubs,GCanas01,GCanas02,QRAC1024,Iemmi,MSolis01,BMarques01,Moha,Mhlam,Vicenzo}.

The experimental setup is presented in Fig.~\ref{fig:setup}. It is composed of two main parts: one for the state preparation and another for performing projective measurements on the prepared system. Each part is controlled by a Field Programmable Gate Array (FPGA) electronics. In the state preparation, a 690 nm single mode laser modulated with an Acousto-Optic Modulator (AOM) and optical attenuators (not shown in Fig. \ref{fig:setup}) prepare weak coherent states with an average number of $0.9$ photons per pulse. This source can be seen as an approximation to a nondeterministic single-photon source, since pulses with a single-photon account for 62.3\% of the generated non-null pulses. Contributions of multiphoton events to the recorded statistics is strongly suppressed by using a detection window much smaller than the optical pulse duration.

To encode the quantum states in the linear transverse momentum of single photons we exploit the pixel-programmability of spatial light modulators (SLMs) \citep{GLima03,TomoMubs}. The state preparation and measurement stages has two fundamental blocks: an \textit{amplitude-modulation only} SLM1 (SLM3), built with two linear polarizers and a liquid crystal display (LCD), and a \textit{phase-modulation only} SLM2 (SLM4), composed of two linear polarizers, two quarter wave plates and an LCD. Each SLM is placed in the image plane of its predecessor. In order to experimentally generate some desired states $\ket{\psi_{x,x_0}}$, a set of $d$ slits with a width of 64 $\mu$m and equal center to center separation are displayed on SLM1 and SLM2. The individual transmittances $t_l$ and phases $\phi_l$ of each slit ``l'' are set to reconstruct the real and imaginary parts of $|\psi_{x,x_0}\ra$. The state of the transmitted photon after the SLM2 is given by $|\psi\rangle = \frac{1}{\sqrt{N}}\sum^{d/2}_{l=-d/2}\sqrt{t_l}e^{i\phi_l}|l\rangle,$ where $N$ is a normalisation constant. The coefficients $t_l$ and the phases $\phi_l$ are independently controlled by the SLM 1 and SLM 2, respectively. To implement the desired measurements at the measurement stage, different amplitude and phase sets of the $d$ slits are used at the SLM3 and SLM4. The transmittances and phases of each set are chosen to post-select for detection one of the required states $\ket{\varphi_{y,b_y}}$. In the final part of the setup, a ``pointlike'' avalanche single-photon detector (APD) with a 10 $\mu$m pinhole is placed at the center of the far field plane of the SLM4. In this case, the probability of single-photon detection $P(x,x_0,y,b)$ is proportional to $|\la\varphi_{y,b}|\psi_{x,x_0}\ra|^2$ \citep{TomoMubs,GCanas01,GCanas02,QRAC1024}. However, since for each $d$ one of the targeted protocol measurements is rank-two projective (see Append.~\ref{App}), we post-process the experimental data to emulate the statistics such a measurement. This is done by suitably relabling the outcomes of the relevant measurements whenever, in the raw data, it is associated to an outcome which never occurs in the desired rank-two projective measurement.

\begin{figure}
	\centering
	\includegraphics[width=\columnwidth]{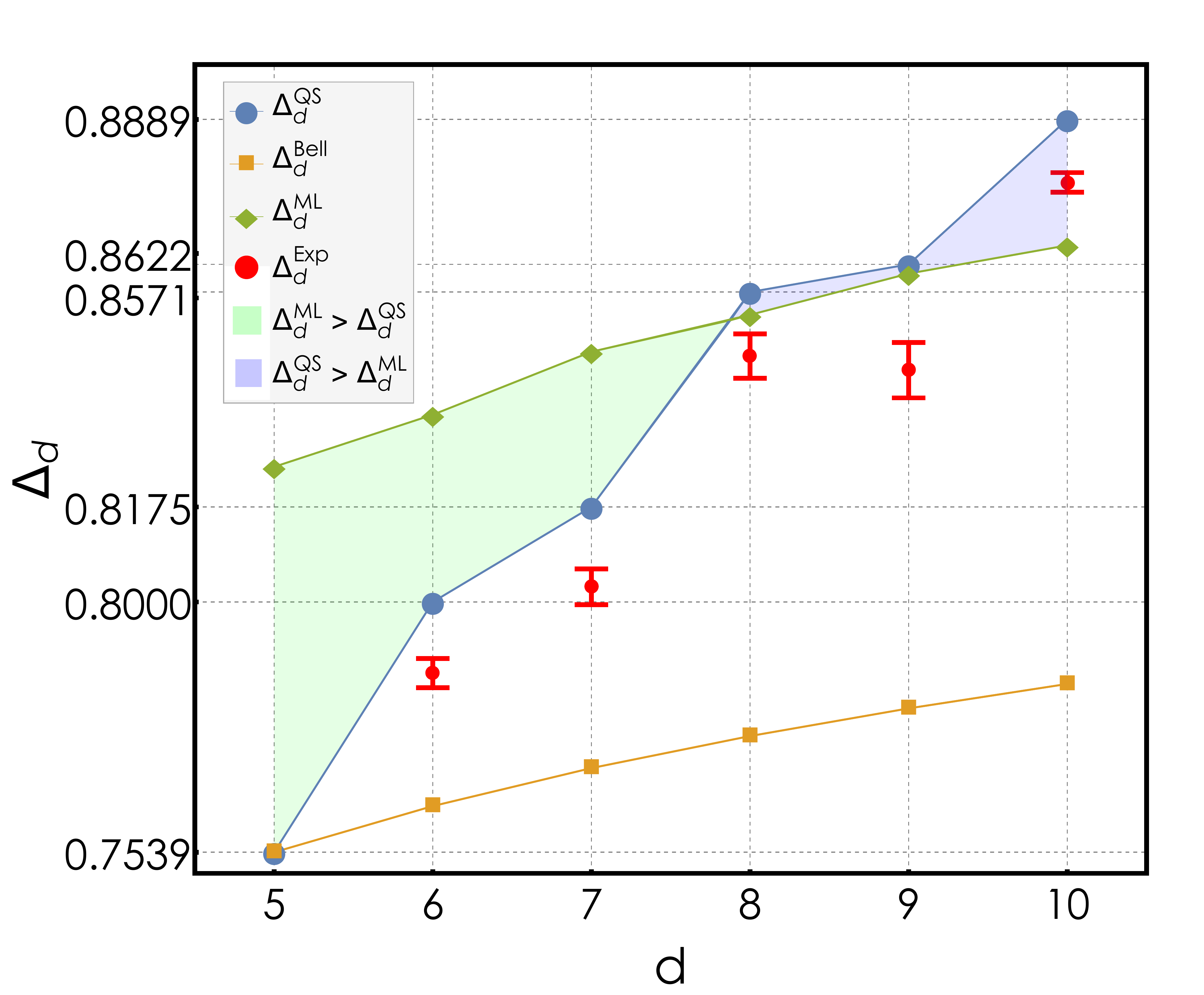}
	\caption{Experimental results. $\Delta^{\text{Exp}}_d$ is represented by red points. The yellow points represent $\Delta_d^{\text{Bell}}$. The blue points are the theoretical predictions of $\Delta_d^{\text{QS}}$. The green points represent $\Delta_d^{\text{ML}}$.}
	\label{fig:results}
\end{figure}

After several rounds of the experiment, an experimental value of $\Delta_d^{\text{QS}}$ is calculated from the acquired data, namely $\Delta_d^{\text{Exp}}$. Since the measurement uncertainty of $\Delta_d^{\text{Exp}}$ decreases with the total number of counts, the repetition of the experimental rounds for each dimension were chosen such that $\Delta_d^{\text{Exp}}$ violates the bounds for Bell inequality based strategies with at least six standard deviations for each $d$ considered. Hence, any explanation
in terms of an arbitrary entangled quantum system is excluded by at least 6 standard deviations,
which corresponds to a p-value of $1\times10^{-9}$.

For $d=6,...,10$, we obtain the results
\begin{align}\nonumber
& \Delta_6^{\text{Exp}}=0.7893\pm 0.0026 & \Delta_7^{\text{Exp}}=0.8082\pm 0.0034 \\\nonumber
& \Delta_8^{\text{Exp}}=0.8453\pm 0.0041 & \Delta_9^{\text{Exp}}=0.8427\pm 0.0051 \\
&  \Delta_{10}^{\text{Exp}}=0.8773\pm 0.0018.
\end{align} In Figure \ref{fig:results}, we compare the experimental results to the theoretical predictions for quantum communication, as well as with the limitations of both quantum and macroscopically local Bell correlations. The results are in good agreement with the theoretical predictions, surpassing the values associated to the maximal violation of the CGLMP inequalities. In the particular, in the case of  $d=10$, the results also surpass the limitations of the post-quantum Bell correlations obeying only macroscopic locality.

Finally, we revisit the previously numerically evidenced hypothesis of rank-one projective measurements being sub-optimal. Focusing on the case of $d=6$, we have considered whether the experimental data can be reproduced by some quantum communication strategy utilising only such measurements. To this end, we have used an intermediate level \cite{comment1} of the hierarchy of dimensionally bounded quantum correlations \cite{NV}, and additionally imposed upper and lower bounds on the particular probabilities measured in the lab corresponding to $(x_0,x)=(4,0)$ and $y=0$. In order to respect the errors of the measured probabilities, they was constrained to an interval twice larger than the experimental errors of each measurement outcome. In this manner, we have obtained the bound $0.7830$ on $\Delta_6^{\text{QS}}$ which is smaller than the experimentally measured value. This demonstrates that under the assumption of a six dimensional Hilbert space, there exists no quantum communication strategy based on rank-one projective measurements which can reproduce the experimental results.

\textit{Conclusion.---} We have theoretically and experimentally studied the efficiency of high-dimensional quantum communication in a family of CCPs, as opposed to classical communication assisted by nonlocal correlations violating the facet Bell inequality to which the CCPs were originally tailored. We demonstrated significant advantages of quantum communication which increase with Hilbert space dimension, and showed that they stem from uncommon types of quantum states and measurements. Our work shows the usefulness and strength of quantum correlations generated via the communication of a high-dimensional quantum system, and the practicality of experimentally realising them.

\begin{acknowledgments}
This work was supported Fondecyt~1160400, Fondecyt~11150324 and Millennium Institute for Research in Optics, MIRO. D.M. acknowledges support from CONICYT Doctorado Nacional 2116050. A.~T.~ acknowledges support from the Swiss national science foundation (Starting grant DIAQ, NCCR-QSIT). B.~M.~ was supported by FAPESP (project 2014/27223-2). A.~T.~ and B.~M.~ thank G.~L.~ and G.~C.~ for their hospitality in Concepci\'on.
\end{acknowledgments}

\onecolumngrid
\appendix

%%%%%%%%%%%%%%%%%%%%%%%%%%%%%%%%%%%%%%%%%%%%%%%%%%%%%%%%%%%%%%%%%%%%%

\newpage

\section{Numerical findings for states and measurements}\label{App}
Since the states and measurements used in the main text were obtained numerically, we here state them explicitly by associating them ("$\sim$") to either rank-one or rank-two projectors as specified below. Do notice that some states for a given dimension are identical, and that some measurement operators are zero-projectors.

\subsection{States}
For dimension $6$:
\begin{tiny}
\begin{eqnarray}
\rho_{0,0}\sim \begin{pmatrix}-0.41\\
-0.38\\
0.45\\
0.05\\
0.07\\
0.69\\
\end{pmatrix}
,\rho_{1,0}\sim \begin{pmatrix}-0.41\\
-0.38\\
0.45\\
0.05\\
0.07\\
0.69\\
\end{pmatrix}
, \rho_{2,0}\sim \begin{pmatrix}0.01\\
-0.24\\
-0.82\\
-0.24\\
-0.16\\
0.44\\
\end{pmatrix}
, \rho_{3,0}\sim \begin{pmatrix}0.01\\
-0.24\\
-0.82\\
-0.24\\
-0.16\\
0.44\\
\end{pmatrix}
, \rho_{4,0}\sim \begin{pmatrix}0.45\\
0.07\\
0.05\\
0.64\\
-0.56\\
0.27\\
\end{pmatrix}
, \rho_{5,0}\sim 
\begin{pmatrix}0.45\\
0.07\\
0.05\\
0.64\\
-0.56\\
0.27\\
\end{pmatrix}
\\,
\rho_{0,1}\sim \begin{pmatrix}0.74\\
0.09\\
0.17\\
-0.29\\
0.47\\
0.35\\
\end{pmatrix},
\rho_{1,1}\sim\begin{pmatrix}0.74\\
0.09\\
0.17\\
-0.29\\
0.47\\
0.35\\
\end{pmatrix}, 
\rho_{2,1}\sim \begin{pmatrix}-0.26\\
0.57\\
-0.27\\
0.47\\
0.50\\
0.25\\
\end{pmatrix}
,
\rho_{3,1}\sim \begin{pmatrix}-0.26\\
0.57\\
-0.27\\
0.47\\
0.50\\
0.25\\
\end{pmatrix}
,
\rho_{4,1}\sim \begin{pmatrix}-0.15\\
0.68\\
0.17\\
-0.48\\
-0.44\\
0.26\\
\end{pmatrix}
,
\rho_{5,1}\sim \begin{pmatrix}-0.15\\
0.68\\
0.17\\
-0.48\\
-0.44\\
0.26\\
\end{pmatrix}.
\end{eqnarray}	
\end{tiny}

For dimension $7$:
\begin{tiny}
\begin{eqnarray}
\rho_{0,0}\sim	\begin{pmatrix}
		-0.00+0.39i\\
		-0.12+0.23i\\
		-0.20-0.30i\\
		0.02-0.20i\\
		-0.12+0.43i\\
		0.17-0.16i\\
		0.60\\
	\end{pmatrix}
, \rho_{1,0}\sim \begin{pmatrix}
-0.00+0.39i\\
-0.12+0.23i\\
-0.20-0.30i\\
0.02-0.20i\\
-0.12+0.43i\\
0.17-0.16i\\
0.60\\
\end{pmatrix}
, \rho_{2,0}\sim \begin{pmatrix}
0.18+0.04i\\
0.13-0.06i\\
0.29-0.13i\\
-0.20-0.60i\\
0.05-0.33i\\
-0.32+0.38i\\
0.30\\
\end{pmatrix}
, \rho_{3,0}\sim \begin{pmatrix}
0.18+0.04i\\
0.13-0.06i\\
0.29-0.13i\\
-0.20-0.60i\\
0.05-0.33i\\
-0.32+0.38i\\
0.30\\
\end{pmatrix}
\\
, \rho_{4,0}\sim \begin{pmatrix}
-0.35-0.40i\\
0.09-0.28i\\
-0.05+0.57i\\
-0.15-0.26i\\
-0.10+0.32i\\
0.06-0.03i\\
0.31\\
\end{pmatrix}
,\rho_{5,0}\sim \begin{pmatrix}
-0.15-0.11i\\
0.51+0.14i\\
0.14-0.03i\\
0.20+0.44i\\
0.26-0.20i\\
0.21+0.20i\\
0.49\\
\end{pmatrix}
, \rho_{6,0}\sim \begin{pmatrix}
-0.15-0.11i\\
0.51+0.14i\\
0.14-0.03i\\
0.20+0.44i\\
0.26-0.20i\\
0.21+0.20i\\
0.49\\
\end{pmatrix}
\\
\rho_{0,1}\sim \begin{pmatrix}0.03-0.09i\\
-0.54+0.17i\\
-0.16+0.26i\\
-0.04-0.03i\\
0.12-0.61i\\
0.24-0.22i\\
0.28\\
\end{pmatrix}
,
\rho_{1,1}\sim\begin{pmatrix}0.41+0.28i\\
-0.19-0.35i\\
0.04+0.35i\\
0.09+0.42i\\
-0.29+0.04i\\
-0.28+0.21i\\
0.28\\
\end{pmatrix}
,
\rho_{2,1}\sim\begin{pmatrix}0.41+0.28i\\
-0.19-0.35i\\
0.04+0.35i\\
0.09+0.42i\\
-0.29+0.04i\\
-0.28+0.21i\\
0.28\\
\end{pmatrix}
,
\rho_{3,1}\sim\begin{pmatrix}0.16-0.25i\\
-0.17-0.02i\\
-0.05-0.68i\\
0.09+0.30i\\
0.05-0.16i\\
-0.47-0.27i\\
0.06\\
\end{pmatrix}
\\
,
\rho_{4,1}\sim\begin{pmatrix}0.16-0.25i\\
-0.17-0.02i\\
-0.05-0.68i\\
0.09+0.30i\\
0.05-0.16i\\
-0.47-0.27i\\
0.06\\
\end{pmatrix}
,
\rho_{5,1}\sim\begin{pmatrix}-0.16-0.61i\\
-0.08-0.33i\\
0.05-0.11i\\
-0.13+0.06i\\
-0.01+0.17i\\
-0.36-0.39i\\
0.37\\
\end{pmatrix}
,
\rho_{6,1}\sim\begin{pmatrix}0.03-0.09i\\
-0.54+0.17i\\
-0.16+0.26i\\
-0.04-0.03i\\
0.12-0.61i\\
0.24-0.22i\\
0.28\\
\end{pmatrix}
.
\end{eqnarray}
\end{tiny}

For dimension 8:
\begin{tiny}
\begin{eqnarray}
\rho_{0,0}\sim	\begin{pmatrix}-0.12+0.24i\\
-0.03+0.08i\\
-0.17-0.02i\\
0.04+0.03i\\
0.06+0.59i\\
-0.15-0.66i\\
0.01-0.29i\\
0.03\\
\end{pmatrix}
,
\rho_{1,0}\sim	\begin{pmatrix}-0.12+0.24i\\
-0.03+0.08i\\
-0.17-0.02i\\
0.04+0.03i\\
0.06+0.59i\\
-0.15-0.66i\\
0.01-0.29i\\
0.03\\
\end{pmatrix},
\rho_{2,0}\sim \begin{pmatrix}0.34-0.05i\\
-0.11-0.45i\\
-0.02-0.15i\\
0.22+0.34i\\
-0.24-0.20i\\
-0.20-0.31i\\
-0.11+0.19i\\
0.44\\
\end{pmatrix}
,	
\rho_{3,0}\sim \begin{pmatrix}0.34-0.05i\\
-0.11-0.45i\\
-0.02-0.15i\\
0.22+0.34i\\
-0.24-0.20i\\
-0.20-0.31i\\
-0.11+0.19i\\
0.44\\
\end{pmatrix}
\\
,	
\rho_{4,0}\sim	\begin{pmatrix}0.20+0.15i\\
-0.14+0.43i\\
0.16-0.27i\\
-0.12+0.22i\\
-0.11+0.06i\\
-0.03+0.33i\\
0.06-0.48i\\
0.47\\
\end{pmatrix}
,
\rho_{5,0}\sim	\begin{pmatrix}0.20+0.15i\\
-0.14+0.43i\\
0.16-0.27i\\
-0.12+0.22i\\
-0.11+0.06i\\
-0.03+0.33i\\
0.06-0.48i\\
0.47\\
\end{pmatrix}
\rho_{6,0}\sim\begin{pmatrix}-0.43+0.20i\\
0.49+0.07i\\
0.21+0.16i\\
0.50-0.06i\\
-0.09-0.03i\\
-0.13+0.08i\\
0.18+0.08i\\
0.36\\
\end{pmatrix}
,
\rho_{7,0}\sim	\begin{pmatrix}-0.43+0.20i\\
0.49+0.07i\\
0.21+0.16i\\
0.50-0.06i\\
-0.09-0.03i\\
-0.13+0.08i\\
0.18+0.08i\\
0.36\\
\end{pmatrix}\\
\rho_{0,1} \sim \begin{pmatrix}-0.39-0.13i\\
-0.03-0.25i\\
-0.23+0.32i\\
-0.30-0.14i\\
-0.31-0.10i\\
-0.15+0.10i\\
-0.42-0.37i\\
0.22\\
\end{pmatrix}
,
\rho_{1,1} \sim \begin{pmatrix}-0.02+0.40i\\
0.03+0.06i\\
-0.76-0.04i\\
-0.17+0.05i\\
0.01-0.03i\\
0.01+0.19i\\
0.23+0.31i\\
0.18\\
\end{pmatrix}
,
\rho_{2,1} \sim \begin{pmatrix}-0.02+0.40i\\
0.03+0.06i\\
-0.76-0.04i\\
-0.17+0.05i\\
0.01-0.03i\\
0.01+0.19i\\
0.23+0.31i\\
0.18\\
\end{pmatrix}
,
\rho_{3,1} \sim \begin{pmatrix}0.02-0.13i\\
-0.23+0.16i\\
0.06+0.15i\\
-0.07-0.45i\\
0.37-0.06i\\
0.34-0.23i\\
-0.09+0.20i\\
0.56\\
\end{pmatrix}
\\
,
\rho_{4,1} \sim \begin{pmatrix}0.02-0.13i\\
-0.23+0.16i\\
0.06+0.15i\\
-0.07-0.45i\\
0.37-0.06i\\
0.34-0.23i\\
-0.09+0.20i\\
0.56\\
\end{pmatrix}
,
\rho_{5,1} \sim \begin{pmatrix}-0.04-0.42i\\
0.30-0.30i\\
0.09-0.05i\\
-0.37+0.16i\\
0.19+0.50i\\
-0.06+0.19i\\
0.26+0.13i\\
0.24\\
\end{pmatrix}
,
\rho_{6,1} \sim \begin{pmatrix}-0.04-0.42i\\
0.30-0.30i\\
0.09-0.05i\\
-0.37+0.16i\\
0.19+0.50i\\
-0.06+0.19i\\
0.26+0.13i\\
0.24\\
\end{pmatrix}
,
\rho_{7,1} \sim \begin{pmatrix}-0.39-0.13i\\
-0.03-0.25i\\
-0.23+0.32i\\
-0.30-0.14i\\
-0.31-0.10i\\
-0.15+0.10i\\
-0.42-0.37i\\
0.22\\
\end{pmatrix}.
\end{eqnarray}
\end{tiny}

For dimension 9:
\begin{tiny}
\begin{eqnarray}
\rho_{0,0}\sim	\begin{pmatrix}-0.10-0.34i\\
-0.25+0.09i\\
-0.20+0.16i\\
-0.46-0.02i\\
0.21-0.02i\\
-0.38+0.16i\\
0.15-0.26i\\
-0.27-0.23i\\
0.29\\
\end{pmatrix}
,
\rho_{1,0}\sim\begin{pmatrix}-0.10-0.34i\\
-0.25+0.09i\\
-0.20+0.16i\\
-0.46-0.02i\\
0.21-0.02i\\
-0.38+0.16i\\
0.15-0.26i\\
-0.27-0.23i\\
0.29\\
\end{pmatrix},
\rho_{2,0}\sim\begin{pmatrix}0.28+0.03i\\
0.50+0.51i\\
-0.06+0.30i\\
-0.05-0.31i\\
-0.00+0.13i\\
0.12-0.19i\\
-0.27-0.06i\\
-0.15-0.09i\\
0.22\\
\end{pmatrix},
\rho_{3,0}\sim \begin{pmatrix}0.28+0.03i\\
0.50+0.51i\\
-0.06+0.30i\\
-0.05-0.31i\\
-0.00+0.13i\\
0.12-0.19i\\
-0.27-0.06i\\
-0.15-0.09i\\
0.22\\
\end{pmatrix}
,
\rho_{4,0}\sim\begin{pmatrix}-0.11-0.01i\\
0.19+0.12i\\
0.48-0.15i\\
0.21-0.16i\\
-0.23-0.03i\\
-0.42+0.37i\\
-0.01-0.13i\\
0.34-0.28i\\
0.20\\
\end{pmatrix}
\\
,
\rho_{5,0}\sim\begin{pmatrix}-0.11-0.01i\\
0.19+0.12i\\
0.48-0.15i\\
0.21-0.16i\\
-0.23-0.03i\\
-0.42+0.37i\\
-0.01-0.13i\\
0.34-0.28i\\
0.20\\
\end{pmatrix},
\rho_{6,0}\sim\begin{pmatrix}0.07+0.30i\\
-0.17-0.20i\\
-0.07+0.40i\\
0.32+0.08i\\
0.25+0.20i\\
-0.22+0.22i\\
-0.22+0.46i\\
-0.13-0.21i\\
0.19\\
\end{pmatrix}
,
\rho_{7,0}\sim\begin{pmatrix}0.17+0.10i\\
-0.23-0.21i\\
0.06+0.01i\\
-0.26+0.02i\\
-0.22-0.36i\\
0.29-0.07i\\
-0.23+0.06i\\
0.26+0.01i\\
0.63\\
\end{pmatrix}
,
\rho_{8,0}\sim\begin{pmatrix}0.17+0.10i\\
-0.23-0.21i\\
0.06+0.01i\\
-0.26+0.02i\\
-0.22-0.36i\\
0.29-0.07i\\
-0.23+0.06i\\
0.26+0.01i\\
0.63\\
\end{pmatrix}\\
, 
\rho_{0,1}\sim \begin{pmatrix}-0.06-0.33i\\
-0.13-0.17i\\
0.10+0.16i\\
0.22-0.09i\\
-0.12+0.63i\\
0.21+0.11i\\
0.05-0.23i\\
0.03+0.37i\\
0.30\\
\end{pmatrix}
,
\rho_{1,1}\sim \begin{pmatrix}-0.06-0.33i\\
-0.13-0.17i\\
0.10+0.16i\\
0.22-0.09i\\
-0.12+0.63i\\
0.21+0.11i\\
0.05-0.23i\\
0.03+0.37i\\
0.30\\
\end{pmatrix}
,
\rho_{2,1}\sim \begin{pmatrix}-0.30-0.50i\\
0.22-0.03i\\
-0.31+0.20i\\
0.30-0.01i\\
0.05-0.28i\\
0.32+0.10i\\
0.13+0.19i\\
0.23-0.29i\\
0.04\\
\end{pmatrix},
\rho_{3,1}\sim \begin{pmatrix}-0.30-0.50i\\
0.22-0.03i\\
-0.31+0.20i\\
0.30-0.01i\\
0.05-0.28i\\
0.32+0.10i\\
0.13+0.19i\\
0.23-0.29i\\
0.04\\
\end{pmatrix}
,
\rho_{4,1}\sim\begin{pmatrix}-0.20+0.37i\\
0.29+0.16i\\
-0.19-0.07i\\
-0.21+0.17i\\
0.05+0.21i\\
0.01+0.07i\\
0.57+0.23i\\
0.21+0.15i\\
0.32\\
\end{pmatrix}
\\
, 
\rho_{5,1}\sim \begin{pmatrix}-0.20+0.37i\\
0.29+0.16i\\
-0.19-0.07i\\
-0.21+0.17i\\
0.05+0.21i\\
0.01+0.07i\\
0.57+0.23i\\
0.21+0.15i\\
0.32\\
\end{pmatrix}
,
\rho_{6,1}\sim \begin{pmatrix}-0.15-0.13i\\
0.17+0.05i\\
-0.05-0.57i\\
0.32+0.10i\\
0.07-0.19i\\
-0.07-0.38i\\
0.03-0.24i\\
-0.30+0.14i\\
0.35\\
\end{pmatrix}
,
\rho_{7,1}\sim \begin{pmatrix}-0.03+0.05i\\
0.01-0.04i\\
0.03-0.29i\\
0.54+0.12i\\
0.30-0.09i\\
-0.14-0.20i\\
-0.02+0.12i\\
-0.44-0.01i\\
0.48\\
\end{pmatrix}
,
\rho_{8,1}\sim \begin{pmatrix}-0.03+0.05i\\
0.01-0.04i\\
0.03-0.29i\\
0.54+0.12i\\
0.30-0.09i\\
-0.14-0.20i\\
-0.02+0.12i\\
-0.44-0.01i\\
0.48\\
\end{pmatrix}.
\end{eqnarray}
\end{tiny}

For dimension 10:
\begin{tiny}
\begin{eqnarray}
\rho_{0,0}\sim\begin{pmatrix}-0.23+0.24i\\
0.03+0.24i\\
0.38-0.00i\\
0.12-0.48i\\
-0.11-0.04i\\
-0.02-0.07i\\
0.09+0.03i\\
0.15-0.59i\\
0.04+0.07i\\
0.20\\
\end{pmatrix}
,
\rho_{1,0}\sim\begin{pmatrix}-0.23+0.24i\\
0.03+0.24i\\
0.38-0.00i\\
0.12-0.48i\\
-0.11-0.04i\\
-0.02-0.07i\\
0.09+0.03i\\
0.15-0.59i\\
0.04+0.07i\\
0.20\\
\end{pmatrix}
,
\rho_{2,0}\sim\begin{pmatrix}0.05-0.06i\\
0.19+0.33i\\
-0.25-0.24i\\
0.35+0.16i\\
0.11+0.14i\\
-0.14+0.15i\\
-0.11+0.29i\\
-0.16-0.04i\\
-0.26+0.34i\\
0.44\\
\end{pmatrix}
,
\rho_{3,0}\sim\begin{pmatrix}0.05-0.06i\\
0.19+0.33i\\
-0.25-0.24i\\
0.35+0.16i\\
0.11+0.14i\\
-0.14+0.15i\\
-0.11+0.29i\\
-0.16-0.04i\\
-0.26+0.34i\\
0.44\\
\end{pmatrix}
,
\rho_{4,0}\sim\begin{pmatrix}-0.35-0.43i\\
0.04+0.02i\\
-0.25-0.01i\\
-0.22-0.07i\\
-0.34+0.01i\\
0.10+0.20i\\
0.15-0.20i\\
-0.01-0.03i\\
-0.05-0.37i\\
0.45\\
\end{pmatrix}
\\
,
\rho_{5,0}\sim\begin{pmatrix}-0.35-0.43i\\
0.04+0.02i\\
-0.25-0.01i\\
-0.22-0.07i\\
-0.34+0.01i\\
0.10+0.20i\\
0.15-0.20i\\
-0.01-0.03i\\
-0.05-0.37i\\
0.45\\
\end{pmatrix}
,
\rho_{6,0}\sim\begin{pmatrix}0.01+0.19i\\
0.23+0.46i\\
0.31+0.02i\\
-0.19+0.36i\\
-0.13-0.30i\\
0.41+0.05i\\
0.21+0.16i\\
0.06+0.26i\\
0.14-0.00i\\
0.10\\
\end{pmatrix}
,
\rho_{7,0}\sim\begin{pmatrix}0.01+0.19i\\
0.23+0.46i\\
0.31+0.02i\\
-0.19+0.36i\\
-0.13-0.30i\\
0.41+0.05i\\
0.21+0.16i\\
0.06+0.26i\\
0.14-0.00i\\
0.10\\
\end{pmatrix}
,
\rho_{8,0}\sim\begin{pmatrix}-0.03+0.42i\\
-0.16-0.32i\\
-0.05-0.14i\\
0.19+0.03i\\
0.29-0.44i\\
-0.03+0.10i\\
0.31-0.02i\\
-0.07+0.06i\\
-0.29-0.31i\\
0.27\\
\end{pmatrix}
,
\rho_{9,0}\sim\begin{pmatrix}-0.03+0.42i\\
-0.16-0.32i\\
-0.05-0.14i\\
0.19+0.03i\\
0.29-0.44i\\
-0.03+0.10i\\
0.31-0.02i\\
-0.07+0.06i\\
-0.29-0.31i\\
0.27\\
\end{pmatrix}
\\
,
\rho_{0,1}\sim\begin{pmatrix}0.25+0.05i\\
-0.22+0.16i\\
0.08+0.23i\\
0.20-0.06i\\
0.22+0.22i\\
0.14-0.30i\\
-0.09-0.32i\\
-0.22+0.15i\\
0.35-0.17i\\
0.46\\
\end{pmatrix}
,
\rho_{1,1}\sim\begin{pmatrix}0.25+0.05i\\
-0.22+0.16i\\
0.08+0.23i\\
0.20-0.06i\\
0.22+0.22i\\
0.14-0.30i\\
-0.09-0.32i\\
-0.22+0.15i\\
0.35-0.17i\\
0.46\\
\end{pmatrix}
,
\rho_{2,1}\sim\begin{pmatrix}-0.11+0.20i\\
-0.11-0.22i\\
0.29-0.23i\\
-0.10-0.13i\\
-0.22+0.26i\\
-0.32+0.19i\\
0.01-0.01i\\
0.23+0.51i\\
0.15+0.23i\\
0.27\\
\end{pmatrix}
,
\rho_{3,1}\sim\begin{pmatrix}-0.11+0.20i\\
-0.11-0.22i\\
0.29-0.23i\\
-0.10-0.13i\\
-0.22+0.26i\\
-0.32+0.19i\\
0.01-0.01i\\
0.23+0.51i\\
0.15+0.23i\\
0.27\\
\end{pmatrix}
,
\rho_{4,1}\sim\begin{pmatrix}-0.02-0.12i\\
0.18-0.11i\\
0.05+0.12i\\
-0.30+0.31i\\
0.28-0.29i\\
-0.28-0.05i\\
-0.02-0.44i\\
0.19-0.22i\\
0.07+0.40i\\
0.24\\
\end{pmatrix}
\\
,
\rho_{5,1}\sim\begin{pmatrix}-0.02-0.12i\\
0.18-0.11i\\
0.05+0.12i\\
-0.30+0.31i\\
0.28-0.29i\\
-0.28-0.05i\\
-0.02-0.44i\\
0.19-0.22i\\
0.07+0.40i\\
0.24\\
\end{pmatrix}
,
\rho_{6,1}\sim\begin{pmatrix}0.23+0.33i\\
0.16+0.14i\\
-0.24+0.50i\\
-0.06+0.12i\\
-0.25+0.08i\\
-0.52+0.09i\\
0.10+0.14i\\
0.10-0.07i\\
0.07-0.27i\\
0.04\\
\end{pmatrix}
,
\rho_{7,1}\sim\begin{pmatrix}0.23+0.33i\\
0.16+0.14i\\
-0.24+0.50i\\
-0.06+0.12i\\
-0.25+0.08i\\
-0.52+0.09i\\
0.10+0.14i\\
0.10-0.07i\\
0.07-0.27i\\
0.04\\
\end{pmatrix}
,
\rho_{8,1}\sim\begin{pmatrix}0.26-0.04i\\
-0.01-0.43i\\
0.03+0.15i\\
-0.28-0.02i\\
-0.09-0.05i\\
0.26-0.22i\\
-0.26+0.51i\\
0.13-0.13i\\
-0.07+0.05i\\
0.37\\
\end{pmatrix},
\rho_{9,1}\sim\begin{pmatrix}0.26-0.04i\\
-0.01-0.43i\\
0.03+0.15i\\
-0.28-0.02i\\
-0.09-0.05i\\
0.26-0.22i\\
-0.26+0.51i\\
0.13-0.13i\\
-0.07+0.05i\\
0.37\\
\end{pmatrix}.
\end{eqnarray}
\end{tiny}

\subsection{Measurements}
For dimension $6$:
\begin{tiny}
\begin{eqnarray}
M_{0}^0\sim 
	\begin{pmatrix}
		-0.41\\
		-0.38\\
		0.45\\
		0.05\\
		0.07\\
		0.69\\
	\end{pmatrix}
, M_{0}^1\sim \begin{tiny}\begin{pmatrix}
0.74\\
0.09\\
0.17\\
-0.29\\
0.47\\
0.35\\
\end{pmatrix}
\end{tiny}
,M_{0}^2\sim \begin{tiny}\begin{pmatrix}
0.01\\
-0.24\\
-0.82\\
-0.24\\
-0.16\\
0.44\\
\end{pmatrix}
\end{tiny}
, M_{0}^3\sim=\begin{tiny}\begin{pmatrix}
-0.26\\
0.57\\
-0.27\\
0.47\\
0.50\\
0.25\\
\end{pmatrix}
\end{tiny}
, M_{0}^4 \sim \begin{tiny}\begin{pmatrix}
0.45\\
0.07\\
0.05\\
0.64\\
-0.56\\
0.27\\
\end{pmatrix}
\end{tiny}
, M_{0}^5\sim \begin{tiny}\begin{pmatrix}
-0.15\\
0.68\\
0.17\\
-0.48\\
-0.44\\
0.26\\
\end{pmatrix}
\end{tiny}
\end{eqnarray}
\end{tiny}
and,  $M_{1}^0=M_{1}^2=M_{1}^4=0$, and
\begin{tiny}
\begin{eqnarray}
M_{1}^1\sim \{
\begin{pmatrix}
0.10\\
-0.67\\
0.41\\
-0.42\\
-0.45\\
-0.00\\
\end{pmatrix}
,\begin{pmatrix}
-0.47\\
-0.16\\
0.33\\
0.21\\
0.23\\
0.73\\
\end{pmatrix}
\}
, M_{1}^3\sim \{
\begin{pmatrix}
0.13\\
-0.69\\
-0.61\\
0.25\\
0.27\\
0.05\\
\end{pmatrix}
,\begin{pmatrix}
-0.08\\
0.20\\
-0.57\\
-0.47\\
-0.38\\
0.51\\
\end{pmatrix}
\}
, M_{1}^5\sim \{
\begin{pmatrix}
-0.86\\
-0.11\\
-0.17\\
-0.10\\
-0.09\\
-0.44\\
\end{pmatrix}
,\begin{pmatrix}
0.02\\
-0.01\\
0.05\\
-0.69\\
0.72\\
-0.04\\
\end{pmatrix}
\}.
\end{eqnarray}
\end{tiny}

For dimension 7: $M_{0}^1=M_{0}^3=M_{0}^6=0$, and
\begin{tiny}
\begin{eqnarray}
M_{0}^0\sim \{\begin{pmatrix}0.17+0.04i\\
-0.48-0.08i\\
-0.39+0.09i\\
-0.06-0.10i\\
0.51-0.27i\\
0.30-0.12i\\
0.35\\
\end{pmatrix},
\begin{pmatrix}-0.09+0.35i\\
-0.10+0.38i\\
-0.05-0.25i\\
0.03-0.17i\\
-0.38+0.32i\\
0.11-0.21i\\
0.56\\
\end{pmatrix}
\},
M_{0}^2\sim \{\begin{pmatrix}-0.23-0.19i\\
-0.13+0.25i\\
-0.14+0.06i\\
0.42+0.30i\\
0.15+0.36i\\
0.33-0.40i\\
-0.35\\
\end{pmatrix},
\begin{pmatrix}-0.43-0.11i\\
0.28+0.13i\\
-0.24-0.38i\\
-0.53-0.21i\\
0.07-0.19i\\
0.28-0.15i\\
-0.21\\
\end{pmatrix}
\}\\
,
M_{0}^4\sim \begin{pmatrix}-0.34-0.22i\\
0.13-0.18i\\
-0.10+0.68i\\
-0.11-0.32i\\
-0.12+0.29i\\
0.20+0.13i\\
0.20\\
\end{pmatrix},
M_{0}^5\sim \{\begin{pmatrix}-0.10+0.43i\\
-0.38+0.35i\\
-0.21+0.17i\\
-0.13-0.17i\\
0.00+0.10i\\
-0.12+0.43i\\
-0.45\\
\end{pmatrix}
,\begin{pmatrix}0.41+0.15i\\
-0.11-0.33i\\
0.01+0.03i\\
-0.01-0.46i\\
-0.33+0.03i\\
0.23-0.41i\\
-0.38\\
\end{pmatrix}
\},
\end{eqnarray}
\end{tiny}
and,
\begin{tiny}
\begin{eqnarray}
M_{1}^0\sim \begin{pmatrix}0.41+0.28i\\
-0.19-0.35i\\
0.04+0.35i\\
0.09+0.42i\\
-0.29+0.04i\\
-0.28+0.21i\\
0.28\\
\end{pmatrix}
,
M_{1}^1\sim \begin{pmatrix}-0.00+0.39i\\
-0.12+0.23i\\
-0.20-0.30i\\
0.02-0.20i\\
-0.12+0.43i\\
0.17-0.16i\\
0.60\\
\end{pmatrix},
M_{1}^2\sim\begin{pmatrix}0.14-0.29i\\
-0.17-0.06i\\
0.01-0.65i\\
0.05+0.29i\\
0.06-0.13i\\
-0.47-0.33i\\
0.10\\
\end{pmatrix},
M_{1}^3\sim \begin{pmatrix}0.18+0.04i\\
0.13-0.06i\\
0.29-0.13i\\
-0.20-0.60i\\
0.05-0.33i\\
-0.32+0.38i\\
0.30\\
\end{pmatrix}\\
,
M_{1}^4\sim \begin{pmatrix}-0.31-0.57i\\
0.02-0.35i\\
-0.00+0.33i\\
-0.16-0.15i\\
-0.07+0.30i\\
-0.13-0.21i\\
0.39\\
\end{pmatrix},
M_{1}^5\sim \begin{pmatrix}0.03-0.09i\\
-0.54+0.17i\\
-0.16+0.26i\\
-0.04-0.03i\\
0.12-0.61i\\
0.24-0.22i\\
0.28\\
\end{pmatrix},
M_{1}^6\sim \begin{pmatrix}-0.15-0.11i\\
0.51+0.14i\\
0.14-0.03i\\
0.20+0.44i\\
0.26-0.20i\\
0.21+0.20i\\
0.49\\
\end{pmatrix}.
\end{eqnarray}
\end{tiny}

For dimension 8:
$M_{0}^1=M_{0}^3=M_{0}^5=M_{0}^7=0$ and
\begin{tiny}
\begin{eqnarray}
M_{0}^0\sim \{\begin{pmatrix}-0.12-0.16i\\
0.05-0.21i\\
-0.14+0.32i\\
-0.23-0.16i\\
0.03-0.37i\\
-0.37+0.42i\\
-0.45-0.19i\\
0.16\\
\end{pmatrix}
,\begin{pmatrix}-0.44+0.02i\\
-0.09-0.12i\\
-0.22+0.13i\\
-0.19-0.03i\\
-0.45+0.33i\\
0.14-0.40i\\
-0.14-0.37i\\
0.16\\
\end{pmatrix}
\},
M_{0}^2\sim \{\begin{pmatrix}0.50+0.13i\\
0.15-0.27i\\
-0.04+0.55i\\
0.01+0.37i\\
-0.07-0.18i\\
0.17-0.21i\\
0.16-0.12i\\
0.22\\
\end{pmatrix}
,\begin{pmatrix}0.09+0.05i\\
-0.18-0.31i\\
-0.33-0.45i\\
0.14+0.18i\\
-0.21-0.13i\\
-0.28-0.14i\\
-0.10+0.38i\\
0.43\\
\end{pmatrix}
\}\\
,
M_{0}^4\sim \{\begin{pmatrix}0.23-0.13i\\
0.16+0.06i\\
-0.26-0.11i\\
0.41+0.02i\\
0.11+0.31i\\
0.43+0.26i\\
-0.49-0.16i\\
0.16\\
\end{pmatrix}
,\begin{pmatrix}0.10+0.02i\\
-0.31+0.39i\\
0.21-0.03i\\
-0.22-0.21i\\
0.20-0.08i\\
0.15-0.02i\\
0.08-0.12i\\
0.71\\
\end{pmatrix}
\},
M_{0}^6\sim \{\begin{pmatrix}0.29+0.09i\\
-0.60+0.15i\\
-0.23-0.08i\\
-0.12+0.01i\\
-0.00-0.32i\\
0.15-0.19i\\
-0.29-0.16i\\
-0.43\\
\end{pmatrix}
,\begin{pmatrix}0.54-0.16i\\
0.09-0.19i\\
0.02-0.14i\\
-0.56-0.31i\\
-0.17+0.40i\\
-0.04+0.06i\\
-0.02+0.10i\\
-0.08\\
\end{pmatrix}
\},
\end{eqnarray}
\end{tiny}
and
\begin{tiny}
\begin{eqnarray}
M_{1}^0\sim \begin{pmatrix}-0.02+0.40i\\
0.03+0.06i\\
-0.76-0.04i\\
-0.17+0.05i\\
0.01-0.03i\\
0.01+0.19i\\
0.23+0.31i\\
0.18\\
\end{pmatrix}
,
M_{1}^1\sim \begin{pmatrix}-0.12+0.24i\\
-0.03+0.08i\\
-0.17-0.02i\\
0.04+0.03i\\
0.06+0.59i\\
-0.15-0.66i\\
0.01-0.29i\\
0.03\\
\end{pmatrix}
,
M_{1}^2\sim \begin{pmatrix}0.02-0.13i\\
-0.23+0.16i\\
0.06+0.15i\\
-0.07-0.45i\\
0.37-0.06i\\
0.34-0.23i\\
-0.09+0.20i\\
0.56\\
\end{pmatrix}
,
M_{1}^3\sim \begin{pmatrix}0.34-0.05i\\
-0.11-0.45i\\
-0.02-0.15i\\
0.22+0.34i\\
-0.24-0.20i\\
-0.20-0.31i\\
-0.11+0.19i\\
0.44\\
\end{pmatrix}
\\
,
M_{1}^4\sim \begin{pmatrix}-0.04-0.42i\\
0.30-0.30i\\
0.09-0.05i\\
-0.37+0.16i\\
0.19+0.50i\\
-0.06+0.19i\\
0.26+0.13i\\
0.24\\
\end{pmatrix}
,
M_{1}^5\sim \begin{pmatrix}0.20+0.15i\\
-0.14+0.43i\\
0.16-0.27i\\
-0.12+0.22i\\
-0.11+0.06i\\
-0.03+0.33i\\
0.06-0.48i\\
0.47\\
\end{pmatrix}
,
M_{1}^6\sim \begin{pmatrix}-0.39-0.13i\\
-0.03-0.25i\\
-0.23+0.32i\\
-0.30-0.14i\\
-0.31-0.10i\\
-0.15+0.10i\\
-0.42-0.37i\\
0.22\\
\end{pmatrix}
,
M_{1}^7\sim \begin{pmatrix}-0.43+0.20i\\
0.49+0.07i\\
0.21+0.16i\\
0.50-0.06i\\
-0.09-0.03i\\
-0.13+0.08i\\
0.18+0.08i\\
0.36\\
\end{pmatrix}.
\end{eqnarray}
\end{tiny}

For dimension 9:
\begin{tiny}
\begin{eqnarray}
M_{0}^0\sim\begin{pmatrix}-0.10-0.34i\\
-0.25+0.09i\\
-0.20+0.16i\\
-0.46-0.02i\\
0.21-0.02i\\
-0.38+0.16i\\
0.15-0.26i\\
-0.27-0.23i\\
0.29\\
\end{pmatrix}
,
M_{0}^1\sim\begin{pmatrix}-0.06-0.33i\\
-0.13-0.17i\\
0.10+0.16i\\
0.22-0.09i\\
-0.12+0.63i\\
0.21+0.11i\\
0.05-0.23i\\
0.03+0.37i\\
0.30\\
\end{pmatrix}
,
M_{0}^2\sim\begin{pmatrix}0.28+0.03i\\
0.50+0.51i\\
-0.06+0.30i\\
-0.05-0.31i\\
-0.00+0.13i\\
0.12-0.19i\\
-0.27-0.06i\\
-0.15-0.09i\\
0.22\\
\end{pmatrix}
,
M_{0}^3\sim\begin{pmatrix}-0.30-0.50i\\
0.22-0.03i\\
-0.31+0.20i\\
0.30-0.01i\\
0.05-0.28i\\
0.32+0.10i\\
0.13+0.19i\\
0.23-0.29i\\
0.04\\
\end{pmatrix}
,
M_{0}^4\sim\begin{pmatrix}-0.11-0.01i\\
0.19+0.12i\\
0.48-0.15i\\
0.21-0.16i\\
-0.23-0.03i\\
-0.42+0.37i\\
-0.01-0.13i\\
0.34-0.28i\\
0.20\\
\end{pmatrix}
\\,
M_{0}^5\sim\begin{pmatrix}-0.20+0.37i\\
0.29+0.16i\\
-0.19-0.07i\\
-0.21+0.17i\\
0.05+0.21i\\
0.01+0.07i\\
0.57+0.23i\\
0.21+0.15i\\
0.32\\
\end{pmatrix}
,
M_{0}^6\sim\begin{pmatrix}-0.09+0.31i\\
-0.05-0.27i\\
-0.36+0.38i\\
0.16+0.10i\\
0.03+0.28i\\
-0.33+0.15i\\
-0.43+0.27i\\
0.09-0.16i\\
0.07\\
\end{pmatrix}
,
M_{0}^7\sim\begin{pmatrix}0.17+0.10i\\
-0.23-0.21i\\
0.06+0.01i\\
-0.26+0.02i\\
-0.22-0.36i\\
0.29-0.07i\\
-0.23+0.06i\\
0.26+0.01i\\
0.63\\
\end{pmatrix}
,
M_{0}^8\sim\begin{pmatrix}-0.04+0.03i\\
0.02-0.03i\\
0.02-0.32i\\
0.53+0.12i\\
0.29-0.11i\\
-0.14-0.22i\\
-0.01+0.08i\\
-0.44+0.00i\\
0.48\\
\end{pmatrix},
\end{eqnarray}
\end{tiny}
and $M_{1}^0=M_{1}^2=M_{1}^4=M_{1}^7$, and
\begin{tiny}
\begin{eqnarray}
M_{1}^1\sim\{\begin{pmatrix}-0.52-0.28i\\
-0.01-0.13i\\
-0.17+0.38i\\
-0.05-0.24i\\
-0.05-0.19i\\
0.06-0.10i\\
0.29-0.08i\\
-0.17-0.45i\\
0.16\\
\end{pmatrix}
,\begin{pmatrix}0.18-0.30i\\
-0.25+0.19i\\
-0.18-0.03i\\
-0.47+0.14i\\
0.29+0.06i\\
-0.44+0.28i\\
0.00-0.23i\\
-0.18-0.02i\\
0.25\\
\end{pmatrix}
\},
M_{1}^3\sim \{
\begin{pmatrix}0.25-0.09i\\
0.52+0.41i\\
0.03+0.28i\\
-0.11-0.34i\\
-0.01+0.11i\\
0.07-0.21i\\
-0.37+0.04i\\
-0.20-0.05i\\
0.18\\
\end{pmatrix}
,\begin{pmatrix}-0.13+0.41i\\
0.31+0.26i\\
-0.23-0.02i\\
-0.17+0.14i\\
0.05+0.22i\\
0.05+0.05i\\
0.54+0.15i\\
0.20+0.11i\\
0.35\\
\end{pmatrix}
\},
M_{1}^5\sim\{
\begin{pmatrix}-0.07-0.08i\\
-0.06+0.08i\\
-0.22+0.20i\\
0.28+0.09i\\
0.05-0.30i\\
-0.37-0.53i\\
0.04-0.07i\\
0.05+0.52i\\
0.12\\
\end{pmatrix}
,\begin{pmatrix}-0.15-0.08i\\
0.24+0.06i\\
0.26-0.62i\\
0.37-0.00i\\
-0.01-0.10i\\
-0.18-0.02i\\
0.01-0.21i\\
-0.17-0.17i\\
0.42\\
\end{pmatrix}\}\\
, M_{1}^6\sim \begin{pmatrix}0.08+0.27i\\
-0.16-0.17i\\
-0.01+0.29i\\
0.42+0.09i\\
0.32+0.14i\\
-0.20+0.16i\\
-0.16+0.46i\\
-0.25-0.19i\\
0.29\\
\end{pmatrix}
, M_{1}^8\sim \{\begin{pmatrix}0.35+0.14i\\
0.02-0.00i\\
-0.13-0.05i\\
-0.17+0.28i\\
-0.52-0.53i\\
-0.09-0.04i\\
0.04+0.26i\\
-0.14-0.27i\\
0.09\\
\end{pmatrix}
,\begin{pmatrix}0.08-0.07i\\
-0.27-0.26i\\
0.11+0.09i\\
-0.12-0.06i\\
-0.19+0.01i\\
0.37-0.01i\\
-0.20-0.07i\\
0.27+0.21i\\
0.69\\
\end{pmatrix}
\}
\end{eqnarray}
\end{tiny}

For dimension 10:
\begin{tiny}
\begin{eqnarray}
M_{0}^0\sim\begin{pmatrix}-0.23+0.24i\\
0.03+0.24i\\
0.38-0.00i\\
0.12-0.48i\\
-0.11-0.04i\\
-0.02-0.07i\\
0.09+0.03i\\
0.15-0.59i\\
0.04+0.07i\\
0.20\\
\end{pmatrix}
,
M_{0}^1\sim\begin{pmatrix}0.25+0.05i\\
-0.22+0.16i\\
0.08+0.23i\\
0.20-0.06i\\
0.22+0.22i\\
0.14-0.30i\\
-0.09-0.32i\\
-0.22+0.15i\\
0.35-0.17i\\
0.46\\
\end{pmatrix}
,
M_{0}^2\sim\begin{pmatrix}0.05-0.06i\\
0.19+0.33i\\
-0.25-0.24i\\
0.35+0.16i\\
0.11+0.14i\\
-0.14+0.15i\\
-0.11+0.29i\\
-0.16-0.04i\\
-0.26+0.34i\\
0.44\\
\end{pmatrix}
,
M_{0}^3\sim\begin{pmatrix}-0.11+0.20i\\
-0.11-0.22i\\
0.29-0.23i\\
-0.10-0.13i\\
-0.22+0.26i\\
-0.32+0.19i\\
0.01-0.01i\\
0.23+0.51i\\
0.15+0.23i\\
0.27\\
\end{pmatrix}
,
M_{0}^4\sim\begin{pmatrix}-0.35-0.43i\\
0.04+0.02i\\
-0.25-0.01i\\
-0.22-0.07i\\
-0.34+0.01i\\
0.10+0.20i\\
0.15-0.20i\\
-0.01-0.03i\\
-0.05-0.37i\\
0.45\\
\end{pmatrix}
\\
,
M_{0}^5\sim\begin{pmatrix}-0.02-0.12i\\
0.18-0.11i\\
0.05+0.12i\\
-0.30+0.31i\\
0.28-0.29i\\
-0.28-0.05i\\
-0.02-0.44i\\
0.19-0.22i\\
0.07+0.40i\\
0.24\\
\end{pmatrix}
,
M_{0}^6\sim\begin{pmatrix}0.01+0.19i\\
0.23+0.46i\\
0.31+0.02i\\
-0.19+0.36i\\
-0.13-0.30i\\
0.41+0.05i\\
0.21+0.16i\\
0.06+0.26i\\
0.14-0.00i\\
0.10\\
\end{pmatrix}
,
M_{0}^7\sim\begin{pmatrix}0.23+0.33i\\
0.16+0.14i\\
-0.24+0.50i\\
-0.06+0.12i\\
-0.25+0.08i\\
-0.52+0.09i\\
0.10+0.14i\\
0.10-0.07i\\
0.07-0.27i\\
0.04\\
\end{pmatrix}
,
M_{0}^8\sim\begin{pmatrix}-0.03+0.42i\\
-0.16-0.32i\\
-0.05-0.14i\\
0.19+0.03i\\
0.29-0.44i\\
-0.03+0.10i\\
0.31-0.02i\\
-0.07+0.06i\\
-0.29-0.31i\\
0.27\\
\end{pmatrix}
,
M_{0}^9\sim\begin{pmatrix}0.26-0.04i\\
-0.01-0.43i\\
0.03+0.15i\\
-0.28-0.02i\\
-0.09-0.05i\\
0.26-0.22i\\
-0.26+0.51i\\
0.13-0.13i\\
-0.07+0.05i\\
0.37\\
\end{pmatrix}.
\end{eqnarray}
\end{tiny}
, and $M_{1}^0=M_{1}^2=M_{1}^4=M_{1}^6=M_{1}^8=0$ and
\begin{tiny}
\begin{eqnarray}
M_{1}^1\sim\{\begin{pmatrix}-0.14+0.31i\\
0.08-0.13i\\
0.44-0.25i\\
-0.19-0.35i\\
-0.31+0.17i\\
-0.34+0.04i\\
0.05-0.03i\\
-0.15+0.20i\\
0.09+0.23i\\
0.28\\
\end{pmatrix}
,\begin{pmatrix}-0.21+0.08i\\
-0.27+0.14i\\
0.17+0.05i\\
0.29-0.17i\\
0.03+0.08i\\
0.04+0.15i\\
0.02+0.07i\\
0.77-0.19i\\
0.12+0.05i\\
0.18\\
\end{pmatrix}
\}
,
M_{1}^3\sim\{\begin{pmatrix}-0.07+0.08i\\
-0.27-0.33i\\
0.31+0.16i\\
-0.24-0.03i\\
-0.23-0.18i\\
0.11-0.08i\\
-0.07-0.25i\\
0.08-0.03i\\
0.31-0.41i\\
-0.43\\
\end{pmatrix}
,\begin{pmatrix}0.05+0.08i\\
-0.03+0.08i\\
-0.13+0.03i\\
0.08-0.52i\\
-0.06+0.32i\\
0.32-0.08i\\
0.33+0.34i\\
-0.03+0.32i\\
-0.13-0.26i\\
-0.25\\
\end{pmatrix}
\}
,
M_{1}^5\sim\{\begin{pmatrix}0.27+0.60i\\
-0.00+0.11i\\
-0.09+0.18i\\
0.11+0.10i\\
0.12-0.04i\\
-0.36-0.30i\\
-0.13+0.26i\\
0.08+0.02i\\
0.15+0.19i\\
-0.34\\
\end{pmatrix}
,\begin{pmatrix}0.19-0.07i\\
-0.08-0.17i\\
0.51-0.26i\\
0.22-0.03i\\
0.41+0.03i\\
0.33+0.03i\\
-0.10-0.02i\\
-0.09+0.03i\\
-0.12+0.38i\\
-0.30\\
\end{pmatrix}
\}
\\
,
M_{1}^7\sim\{\begin{pmatrix}0.12-0.14i\\
-0.35-0.42i\\
-0.08+0.25i\\
-0.32-0.20i\\
0.10+0.03i\\
-0.00-0.05i\\
-0.30+0.47i\\
-0.05-0.19i\\
-0.11+0.11i\\
0.26\\
\end{pmatrix}
,\begin{pmatrix}0.24+0.13i\\
0.38-0.02i\\
0.22-0.02i\\
-0.15+0.29i\\
-0.26-0.19i\\
0.48-0.23i\\
0.00+0.30i\\
0.24+0.10i\\
0.05-0.04i\\
0.29\\
\end{pmatrix}
\}
,
M_{1}^9\sim\{\begin{pmatrix}-0.28-0.40i\\
0.41+0.14i\\
0.14-0.05i\\
-0.24+0.04i\\
0.00+0.29i\\
-0.18-0.01i\\
-0.26+0.33i\\
0.15-0.07i\\
0.16+0.03i\\
-0.37\\
\end{pmatrix}
,\begin{pmatrix}0.01-0.04i\\
0.03+0.11i\\
0.21+0.14i\\
0.14-0.01i\\
0.48+0.24i\\
-0.03-0.30i\\
-0.15-0.08i\\
-0.17+0.16i\\
0.38-0.39i\\
0.38\\
\end{pmatrix}
\}.
\end{eqnarray}
\end{tiny}


\begin{thebibliography}{99}

\bibitem{KSbook}
E. Kushilevitz, and N. Nisan, Communication complexity. Cambridge University Press, 1997.

\bibitem{BB06}
G. Brassard, H. Buhrman, N. Linden, A. A. M\'ethot, A. Tapp, and F. Unger, 
Limit on Nonlocality in Any World in Which Communication Complexity Is Not Trivial,
\href{https://doi.org/10.1103/PhysRevLett.96.250401}{Phys. Rev. Lett. \textbf{96}, 250401 (2006).}

\bibitem{IC09}
M. Paw\l{}owski, T. Paterek, D. Kaszlikowski, V. Scarani, A. Winter, and M. \.Zukowski,
Information Causality as a Physical Principle,
\href{https://doi.org/10.1038/nature08400}{Nature \textbf{461}, pages 1101 (2009).}

\bibitem{Landauer}
R. Landauer,
The physical nature of information,
\href{https://doi.org/10.1016/0375-9601(96)00453-7}{Phys. Lett. A \textbf{217}, 188 (1996).}

\bibitem{ComplexityReview}
H. Buhrman, R. Cleve, S. Massar, and R. de Wolf,
Nonlocality and communication complexity,
\href{https://doi.org/10.1103/RevModPhys.82.665}{Rev. Mod. Phys. \textbf{82}, 665 (2010).}

\bibitem{BC16}
H. Buhrman, L. Czekaj, A. Grudka, M. Horodecki, P. Horodecki, M. Markiewicz, F. Speelman, and S. Strelchuk,
Quantum communication complexity advantage implies violation of a Bell inequality,
\href{https://doi.org/10.1073/pnas.1507647113}{PNAS  \textbf{22}, 113 (2016).}


\bibitem{BZ04}
C. Brukner, M. \.Zukowski, J-W Pan, and A. Zeilinger,
Bell’s Inequalities and Quantum Communication Complexity,
\href{https://doi.org/10.1103/PhysRevLett.92.127901}{Phys. Rev. Lett. \textbf{92}, 127901 (2004).}



\bibitem{TZ17}
A. Tavakoli and M. \.Zukowski,
Higher-dimensional communication complexity problems: Classical protocols versus quantum ones based on Bell's theorem or prepare-transmit-measure schemes,
\href{https://doi.org/10.1103/PhysRevA.95.042305}{Phys. Rev. A \textbf{95}, 042305 (2017).}


\bibitem{CB97}
R. Cleve, and H. Buhrman,
Substituting quantum entanglement for communication,
\href{https://doi.org/10.1103/PhysRevA.56.1201}{Phys. Rev. A \textbf{56}, 1201 (1997).}

\bibitem{Buhrman2}
H. Buhrman, W. van Dam, P. H{\o}yer, and A. Tapp,
Multiparty Quantum Communication Complexity,
\href{https://doi.org/10.1103/PhysRevA.60.2737}{Phys. Rev. A \textbf{60}, 2737 (1999).}

\bibitem{HD199}
L. Hardy, and W. van Dam,
Quantum Whispers,
\href{https://doi.org/10.1103/PhysRevA.59.2635}{Phys. Rev. A \textbf{59}, 2635 (1999).}

\bibitem{BZZ02}
C. Brukner, M. \.Zukowski, and A, Zeilinger,
Quantum communication complexity protocol with two entangled qutrits,
\href{https://doi.org/10.1103/PhysRevLett.89.197901}{Phys. Rev. Lett. \textbf{89}, 197901 (2002).}

\bibitem{BP03}
C. Brukner, T. Paterek, and M. \.Zukowski,
Quantum communication complexity protocols based on higher-dimensional entangled systems,
\href{https://doi.org/10.1142/S0219749903000395}{Int. J. Quantum Inform. \textbf{01}, 519 (2003).}

\bibitem{EB13}
M. Epping, and C. Brukner,
%Bound entanglement helps to reduce communication complexity
\href{https://doi.org/10.1103/PhysRevA.87.032305}{Phys. Rev. A \textbf{87}, 032305 (2013).}

\bibitem{Bridge}
S. Muhammad, A. Tavakoli, M. Kurant, M. Paw\l{}owski, M. \.Zukowski, and M. Bourennane,
%Quantum Bidding in Bridge
\href{https://doi.org/10.1103/PhysRevX.4.021047}{Phys. Rev. X \textbf{4}, 021047 (2014).}


\bibitem{Holevo}
A. S. Holevo,
Bounds for the quantity of information transmitted by a quantum communication channel,
Problems of Information Transmission. \textbf{9}, 177 (1973).

\bibitem{RAC1}
A. Ambainis, A. Nayak, A. Ta-Shma, U. Vazirani,
Dense quantum coding and a lower bound for 1-way quantum automata,
\href{https://doi.org/10.1145/301250.301347}{Proceedings of the 31st Annual ACM Symposium on Theory of Computing (STOC’99), pp. 376–383, 1999.}

\bibitem{Nayak}
A. Nayak,
Optimal lower bounds for quantum automata and random access codes,
\href{https://doi.org/10.1109/SFFCS.1999.814608}{Proceedings of the 40th IEEE Symposium on Foundations of Computer Science (FOCS’99), pp. 369–376, 1999.}

\bibitem{G02}
E. F. Galv\~{a}o,
Feasible quantum communication complexity protocol,
\href{https://doi.org/10.1103/PhysRevA.65.012318}{Phys. Rev. A \textbf{65}, 012318 (2001).}


\bibitem{RACthesis}
A. Ambainis, D. Leung, L. Mancinska, M. Ozols,
Quantum Random Access Codes with Shared Randomness,
arXiv:0810.2937.

\bibitem{WignerFunction}
A. Casaccino, E. F. Galv\~{a}o, and S. Severini,
Extrema of discrete Wigner functions and applications,
\href{https://doi.org/10.1103/PhysRevA.78.022310}{Phys. Rev. A \textbf{78}, 022310 (2008).}

\bibitem{TS05}
P. Trojek, C. Schmid, M. Bourennane, C. Brukner, M. \.Zukowski, and H. Weinfurter,
Experimental quantum communication complexity,
\href{https://doi.org/10.1103/PhysRevA.72.050305}{Phys. Rev. A \textbf{72}, 050305(R) (2005).}

\bibitem{TH15}
A. Tavakoli, A. Hameedi, B. Marques, and M. Bourennane,
% Quantum Random Access Codes Using Single d-Level Systems
\href{https://doi.org/10.1103/PhysRevLett.114.170502}{Phys. Rev. Lett. \textbf{114}, 170502 (2015).}

\bibitem{GF16}
P. A. Gu\'erin, A. Feix, M. Ara\'ujo, and C. Brukner,
Exponential Communication Complexity Advantage from Quantum Superposition of the Direction of Communication,
\href{https://doi.org/10.1103/PhysRevLett.117.100502}{Phys. Rev. Lett. \textbf{117}, 100502 (2016).}

\bibitem{SE16}
M. Smania, A. M Elhassan, A. Tavakoli, and M. Bourennane,
Experimental quantum multiparty communication protocols,
\href{https://doi.org/10.1038/npjqi.2016.10}{npj Quantum Information \textbf{2}, 16010 (2016).}

\bibitem{Crypto}
A. K. Ekert,
Quantum cryptography based on Bell’s theorem,
\href{https://doi.org/10.1103/PhysRevLett.67.661}{Phys. Rev. Lett. \textbf{67}, 661 (1991).};
C. H. Bennett, G. Brassard, and N. D. Mermin,
Quantum cryptography without Bell’s theorem,
\href{https://doi.org/10.1103/PhysRevLett.68.557}{Phys. Rev. Lett. \textbf{68}, 557 (1992).}

\bibitem{Byzantine}
M. Fitzi, N. Gisin, and U. Maurer,
Quantum Solution to the Byzantine Agreement Problem,
\href{https://doi.org/10.1103/PhysRevLett.87.217901}{Phys. Rev. Lett. \textbf{87}, 217901 (2001).};
A. Tavakoli, A. Cabello, M. \.Zukowski, and M. Bourennane,
Quantum Clock Synchronization with a Single Qudit,
\href{https://doi.org/10.1038/srep07982}{Scientific Reports \textbf{5}, 7982 (2015).}

\bibitem{Secretsharing}
M. \.Zukowski, A. Zeilinger, M. A. Horne, and H. Weinfurter,
%Quest for GHZ States
\href{https://doi.org/10.12693/APhysPolA.93.187}{ Acta Phys. Pol. \textbf{93}, 187 (1998).};
M. Hillery, V. Bu\v{z}ek, and A Berthiaume,
%Quantum secret sharing
\href{https://doi.org/10.1103/PhysRevA.59.1829}{Phys. Rev. A \textbf{59}, 1829 (1999).};
C. Schmid, P. Trojek, M. Bourennane, C. Kurtsiefer, M. \.Zukowski, and H. Weinfurter,
Experimental Single Qubit Quantum Secret Sharing,
\href{https://doi.org/10.1103/PhysRevLett.95.230505}{Phys. Rev. Lett. \textbf{95}, 230505 (2005).}

\bibitem{PW}
M. Paw\l{}owski, A. Winter,
Hyperbits: the information quasiparticles,
\href{https://doi.org/10.1103/PhysRevA.85.022331}{Phys. Rev. A \textbf{85}, 022331 (2012).}

\bibitem{PZ10}
M. Paw\l{}owski, and M. \.Zukowski,
Entanglement-assisted random access codes,
\href{https://doi.org/10.1103/PhysRevA.81.042326}{Phys. Rev. A \textbf{81}, 042326 (2010).}

\bibitem{HS17}
A. Hameedi, D. Saha, P. Mironowicz, M. Paw\l owski, M. Bourennane
Complementarity between entanglement-assisted and quantum distributed random access code,
\href{https://doi.org/10.1103/PhysRevA.95.052345}{Phys. Rev. A \textbf{95}, 052345 (2017).}


\bibitem{TM16}
A. Tavakoli, B. Marques, M. Paw\l owski, and M. Bourennane,
Spatial versus sequential correlations for random access coding,
\href{https://doi.org/10.1103/PhysRevA.93.032336}{Phys. Rev. A \textbf{93}, 032336 (2016).}

\bibitem{magic7}
A. Tavakoli, M. Paw\l owski, M. \.Zukowski, and M. Bourennane,
Dimensional discontinuity in quantum communication complexity at dimension seven,
\href{https://doi.org/10.1103/PhysRevA.95.020302}{Phys. Rev. A \textbf{95}, 020302(R) (2017).}

\bibitem{cglmp02}
D. Collins, N. Gisin, N. Linden, S. Massar, and S. Popescu,
Bell Inequalities for Arbitrarily High-Dimensional Systems,
\href{https://doi.org/10.1103/PhysRevLett.88.040404}{Phys. Rev. Lett. \textbf{88}, 040404 (2002).}

\bibitem{M02}
L. Masanes,
Tight Bell inequality for d-outcome measurements correlations,
Quantum Information and Computation, \textbf{3}, 345 (2002).

\bibitem{ML}
M. Navascu\'es, and H. Wunderlich,
A glance beyond the quantum model,
\href{https://doi.org/10.1098/rspa.2009.0453}{Proc. R. Soc. Lond. A \textbf{466} 881–90 (2009).}


\bibitem{PB11}
M. Paw\l{}owski, and N. Brunner,
Semi-device-independent security of one-way quantum key distribution,
\href{https://doi.org/10.1103/PhysRevA.84.010302}{Phys. Rev. A \textbf{84}, 010302(R) (2011).}


\bibitem{Seesaw1}
R. F. Werner, and M. M. Wolf, Bell inequalities and Entanglement, 
Quantum Inf. Comput. \textbf{1}, 1 (2001).

\bibitem{Seesaw2}
K. F. P\'al, and T. V\'ertesi,
Maximal violation of a bipartite three-setting, two-outcome Bell inequality using infinite-dimensional quantum systems,
\href{https://doi.org/10.1103/PhysRevA.82.022116}{Phys. Rev. A \textbf{82}, 022116 (2010).}

\bibitem{sdp}
L. Vandenberghe, and S. Boyd, semidefinite programming,
\href{https://doi.org/10.1137/1038003}{SIAM Rev. \textbf{38}, 49 (1996).}




\bibitem{ZG08}
S. Zohren, and R. Gill,
Maximal violation of the Collins-Gisin-Linden-Massar-Popescu inequality for infinite dimensional states,
\href{https://doi.org/10.1103/PhysRevLett.100.120406}{Phys. Rev. Lett. \textbf{100}, 120406 (2008).}


\bibitem{NV}
M. Navascu\'es, and T. V\'ertesi,
Bounding the Set of Finite Dimensional Quantum Correlations,
\href{https://doi.org/10.1103/PhysRevLett.115.020501}{Phys. Rev. Lett. \textbf{115}, 020501 (2015).}

\bibitem{symmetrisation}
A. Tavakoli, D. Rosset, and M-O. Renou,
Efficiently bounding finite-dimensional quantum correlations via symmetrisation,
In preparation.

\bibitem{GLima03}
G. Lima, A. Vargas, L. Neves, R. Guzm\'an, and C. Saavedra,
Manipulating spatial qudit states with programmable optical devices
\href{https://doi.org/10.1364/OE.17.010688}{Optics Express 13, \textbf{17}, 10688–10696 (2009).}

\bibitem{TomoMubs}
G. Lima, L. Neves, R. Guzm\'an, E. S. G\'omez, W. A. T. Nogueira, A. Delgado, A. Vargas and C. Saavedra,
Experimental quantum tomography of photonic qudits via mutually unbiased basis,
\href{https://doi.org/10.1364/OE.19.003542}{Optics Express 19, \textbf{4}, 3542-3552 (2011).}

\bibitem{GCanas01}
D. Goyeneche, G. Ca\~nas, S. Etcheverry, E. S. G\'omez, G. B. Xavier, G. Lima, and A. Delgado,
Five Measurement Bases Determine Pure Quantum States on Any Dimension,
\href{https://doi.org/10.1103/PhysRevLett.115.090401}{Phys. Rev. Lett. 115, \textbf{423}, 090401 (2015).}

\bibitem{GCanas02}
G. Ca\~nas, S. Etcheverry, E. S. G\'omez, C. Saavedra, G. B. Xavier, G. Lima, and A. Cabello,
Experimental implementation of an eight-dimensional Kochen-Specker set and observation of its connection with the Greenberger-Horne-Zeilinger theorem,
\href{https://doi.org/10.1103/PhysRevA.90.012119}{Phys. Rev. A \textbf{90}, 012119 (2014).}

\bibitem{QRAC1024}
E. A Aguilar, M. Farkas, D. Mart\'inez, M. Alvarado, J. Cari~ne, G. B. Xavier, J. F. Barra, G. Ca\~nas, M. Paw\l{}owski, and G. Lima,
Certifying an irreducible 1024-dimensional photonic state using refined dimension witnesses,
\href{https://arxiv.org/abs/1710.04601}{Phys. Rev. Lett. \textbf{120}, 230503 (2018).}

\bibitem{Iemmi} Q. P. Stefano, L. Reb\'{o}n, S. Ledesma, and C. Iemmi, Determination of any pure spatial qudits from a minimum number of measurements by phase-stepping interferometry, Phys. Rev. A \textbf{96}, 062328 (2017).

\bibitem{MSolis01}
M. A. Sol\'is-Prosser, M. F. Fernandes, O. Jim\'enez, A. Delgado, and L. Neves
Experimental Minimum-Error Quantum-State Discrimination in High Dimensions
\href{https://doi.org/10.1103/PhysRevLett.118.100501}{Phys. Rev. Lett., \textbf{118}, 100501  (2017).}

\bibitem{BMarques01}
B. Marques, A. A. Matoso, W. M. Pimenta, A. J. Guti\'errez-Esparza, M. F. Santos and S. P\'adua
Experimental simulation of decoherence in photonics qudits
\href{https://doi.org/10.1038/srep16049}{Scientific Reports, \textbf{5}, 16049  (2017).}


\bibitem{Moha} M. Mirhosseini, O. S Mag\~na-Loaiza, M. N. O’Sullivan, B. Rodenburg, M. Malik, M. P. J. Lavery, M. J. Padgett, D. J Gauthier, and R. W Boyd,
High-dimensional quantum cryptography with twisted light,
\href{https://doi.org/10.1088/1367-2630/17/3/033033}{ New J. Phys. \textbf{17}, 033033 (2015).}

\bibitem{Mhlam} M. Mafu, A. Dudley, S. Goyal, D. Giovannini, M. McLaren, M. J. Padgett,  T. Konrad, F. Petruccione, N. Lutkenhaus, and A. Forbes,
Higher-dimensional orbital-angular-momentum-based quantum key distribution with mutually unbiased bases,
\href{https://doi.org/10.1103/PhysRevA.88.032305}{Phys. Rev. A \textbf{88}, 032305, (2013).}

\bibitem{Vicenzo} V. D’Ambrosio, F. Bisesto, F. Sciarrino, J. F. Barra, G. Lima, and A. Cabello,
Device-Independent Certification of High-Dimensional Quantum Systems,
\href{https://doi.org/10.1103/PhysRevLett.112.140503}{Phys. Rev. Lett. \textbf{112}, 140503 (2014).}


\bibitem{moreno}
I. Moreno, P. Vel\'asquez, C. R. Fern\'andez-Pousa, M. M. S\'anchez-L\'opez, and F. Mateos,
Jones matrix method for predicting and optimizing the optical modulation properties of a liquid-crystal display
\href{https://doi.org/10.1063/1.1601688}{Journal of Applied Physics \textbf{94}, 3697 (2003).}




\bibitem{comment1}
The hierarchy level is sometimes refered to as 1+AB+BB and corresponds to monomials of the form $\{\openone, \rho_{x_0,x}, M_y^b, \rho_{x_0,x}M_y^b,M_y^bM_{y'}^{b'}\}$.



\end{thebibliography}
\end{document}